\documentclass[onecolumn]{aastex63}

\usepackage{amsmath}
\usepackage{booktabs}
\usepackage{threeparttable}
\usepackage{tabularx}

\usepackage{listings}

\usepackage{lineno}

\submitjournal{ApJ}

\shorttitle{EoR History from 21 cm--NIRB Cross-correlation}
\shortauthors{Sun et al.}
\graphicspath{{./}{figures/}}

\begin{document}

\title{LIMFAST. III. Timing Cosmic Reionization with the 21 cm and Near-Infrared Backgrounds}

\email{guochao.sun@northwestern.edu}

\author{Guochao Sun}
\affiliation{CIERA and Department of Physics and Astronomy, Northwestern University, 1800 Sherman Ave, Evanston, IL 60201, USA}
\affiliation{California Institute of Technology, 1200 E. California Blvd., Pasadena, CA 91125, USA}

\author{Adam Lidz}
\affiliation{University of Pennsylvania, Department of Physics \& Astronomy, 209 S. 33rd Street, Philadelphia, PA 19104, USA}

\author{Tzu-Ching Chang}
\affiliation{Jet Propulsion Laboratory, California Institute of Technology, 4800 Oak Grove Drive, Pasadena, CA 91109, USA}
\affiliation{California Institute of Technology, 1200 E. California Blvd., Pasadena, CA 91125, USA}

\author{Jordan Mirocha}
\affiliation{Jet Propulsion Laboratory, California Institute of Technology, 4800 Oak Grove Drive, Pasadena, CA 91109, USA}
\affiliation{California Institute of Technology, 1200 E. California Blvd., Pasadena, CA 91125, USA}

\author{Steven~R.~Furlanetto}
\affiliation{Department of Physics and Astronomy, University of California, Los Angeles, CA 90095, USA}

\begin{abstract}
The timeline of cosmic reionization remains uncertain despite sustained efforts to study how the ionizing output of early galaxies shaped the intergalactic medium (IGM). Using the semi-numerical code LIMFAST, we investigate the prospects for timing the reionization process by cross-correlating the 21\,cm signal with the cosmic near-infrared background (NIRB) contributed by galaxies at $z>5$. Tracing opposite phases of the IGM on large scales during reionization, the two signals together serve as a powerful probe for the reionization history. However, because long-wavelength, line-of-sight Fourier modes---the only modes probed by NIRB fluctuations---are contaminated by 21\,cm foregrounds and thus inevitably lost to foreground cleaning or avoidance, a direct cross-correlation of the two signals vanishes. We show that this problem can be circumvented by squaring the foreground-filtered 21 cm signal and cross-correlating the squared field with the NIRB. This statistic is related to the 21\,cm--21\,cm--NIRB cross-bispectrum and encodes valuable information regarding the reionization timeline. Particularly, the 21\,cm$^2$ and NIRB signals are positively correlated during the early phases of reionization and negatively correlated at later stages. We demonstrate that
this behavior is generic across several different reionization models and compare our simulated results with perturbative calculations. We show that this cross-correlation can be detected at high significance by forthcoming 21\,cm and NIRB surveys such as SKA and SPHEREx. Our methodology is more broadly applicable to cross-correlations between line intensity mapping data and 2D tracers of the large-scale structure, including photometric galaxy surveys and CMB lensing mass maps, among others. 
\end{abstract}

\keywords{cosmology: observation -- cosmology: theory -- large-scale structure of universe -- reionization}


\section{Introduction}

Observational progress made over the last two decades, especially measurements of the Gunn-Peterson effect \citep{GunnPeterson1965} in high-redshift quasar spectra \citep{Becker2001,Fan2006,McGreer2011,Becker2015,Zhu2021,Bosman2022} and of the electron scattering optical depth of cosmic microwave background (CMB) photons \citep{Larson2011,Planck2016EoR}, has led to constraints on the history of the Epoch of Reionization (EoR). These include estimates of reionization's endpoint and overall duration, allowing meaningful tests of reionization and high-$z$ galaxy formation models \cite[for a recent view, see][]{GnedinMadau2022}. However, neutral fraction estimates of the intergalactic medium (IGM) from quasar absorption spectra are limited by the immense scattering cross section of photons near the Ly$\alpha$ wavelength with neutral hydrogen that yields a large Gunn-Peterson optical depth even with a small neutral fraction \citep{Peebles1993,Miralda-Escude1998}, whereas the reconstruction of full reionization history from only the integral CMB constraints is likely subject to systematic uncertainties associated with e.g., prior assumptions \citep{MilleaBouchet2018}. While alternative probes such as observations of Ly$\alpha$ emission from star-forming galaxies \citep{McQuinn2007,Stark2010,Dijkstra2014,Schenker2014,Mason2019,Jones2024} and the Ly$\alpha$ damping wing in individual or stacked spectra of high-$z$ galaxies, quasars, or gamma-ray bursts \citep{Miralda-Escude1998,McQuinn2008,Davies2018,Lidz2021,Keating2024,Spina2024,Umeda2024} have been developed to (partially) overcome these shortcomings, it is still desirable to constrain the global reionization timeline by observing the evolution of the neutral IGM directly. 

Intensity mapping (IM) of the 21\,cm spin-flip transition of atomic hydrogen has consequently emerged as a promising technique to directly measure the detailed three-dimensional (3D) distribution of neutral hydrogen and its time evolution during the EoR \citep{Furlanetto2006,PritchardLoeb2012,Mesinger2019}. Statistical analysis of the redshifted 21\,cm line, in particular its spatial fluctuations, can reveal rich physical information on the reionization process, including the global neutral fraction, topological structures at different stages, and the nature of the ionizing sources. Besides probing early galaxies and the IGM, 21\,cm IM of the EoR and the cosmic dark ages also probes cosmology \citep{LiuParsons2016,Kern2017,Hassan2020}, placing useful constraints on e.g., cosmic inflation \citep{Joudaki2011,Munoz2015,Greig2023} and alternative dark matter models \citep{Sitwell2014,Jones2021,FlitterKovetz2022}. One major challenge for studying the EoR with 21\,cm IM is the treatment of bright contaminating foregrounds and instrumental systematics \citep{LiuShaw2020}. Cross-correlating the 21\,cm signal with another complementary tracer is a powerful and reliable means to extract information from noisy 21\,cm data, which makes use of the fact that surveys of different tracers usually have independent, uncorrelated sources of foreground contamination and systematics \citep{FurlanettoLidz2007,Lidz2009,Lidz2011,Ma2018,Beane2019,LaPlante2020,LaPlante2023,Moriwaki2024,McBrideLiu2024}. 

Indeed, in addition to the 21\,cm line, the EoR is being studied with the IM approach in a multitude of tracers, most of which generally trace cosmic star formation in galaxies that create ionized regions in the IGM during the EoR \citep{BernalKovetz2022}. Different types of tracer often probe different source populations (e.g., normal star-forming galaxies, quasars, Population~III stars), thus providing unique and complementary information about the high-$z$ universe. Synergies of these line or continuum IM signals with the 21\,cm line have been extensively investigated ever since the proposal of the 21\,cm line as a probe of cosmic reionization.  Such IM measurements can be conducted in 3D for some redshifted bright emission lines, such as Ly$\alpha$ \citep{Pullen_2014,Heneka_2017,Cox2022,VisbalMcQuinn2023}, \ion{He}{2} \citep{Visbal2015,Parsons2022}, [\ion{C}{2}] \citep{Gong_2012,Silva2015,Dumitru2019,Sun_2021ApJ,FronenbergLiu2024}, [\ion{O}{3}] \citep{Padmanabhan_2022,Padmanabhan_2023}, and CO \citep{Lidz2011,Zhou2021,Breysse_2022}. Alternatively, angular fluctuations of the extragalactic background light (EBL; \citealt{Cooray2016}) in certain photometric bands, such as the cosmic near-infrared background (NIRB; \citealt{Mao2014}) and the cosmic X-ray background (XRB; \citealt{Ma2018XRB}), can be measured.

The cosmic NIRB is a particularly interesting signal because it contains the redshifted rest-UV emission from high-$z$ star-forming galaxies responsible for the ionizing radiation background during the EoR. This high-$z$ component of the NIRB may be extracted using multi-band and multi-scale information \cite[][]{Feng2019} or by cross-correlating with other tracers like galaxies \citep{ChengChang2022}. The redshift-integrated nature of the NIRB makes it a useful probe of cosmic star formation history during the EoR \cite[for a recent review, see][]{Kashlinsky2018}, which motivates explorations of the joint analysis between NIRB and 21\,cm surveys \citep{Fernandez2014,Mao2014}. As demonstrated in \citet{Mao2014}, when deep, wide-area 21\,cm and NIRB surveys are cross-correlated such that the redshift degeneracy of the NIRB is lifted, it is possible to simultaneously constrain both the progress of reionization and physical properties of the early, star-forming galaxies as ionizing sources. While the 21\,cm--NIRB cross-correlation is scientifically interesting, we note that the existing studies of it involve somewhat outdated models of high-$z$ galaxy formation and unrealistic assumptions about 21\,cm foreground cleaning. With new-generation observing facilities such as the Square Kilometer Array \citep[SKA;][]{Mellema2013SKA} for the 21\,cm line and the NASA MIDEX mission Spectro-Photometer for the History of the Universe, Epoch of Reionization and Ices Explorer \citep[SPHEREx;][]{Dore2014} for the NIRB starting operations soon, an updated and more thorough investigation of this useful cross-correlation signal is in need. 

In this work, we extend the semi-numerical code LIMFAST \citep{LIMFAST2023a,LIMFAST2023b} to forward model the NIRB signal contributed by galaxies at $z>5$, in a way fully consistent with calculations of the 21\,cm signal for the investigation of their cross-correlation during the EoR. Though the idea of 21\,cm--NIRB cross-correlation is not new, we introduce two major improvements over previous studies. First, by considering the loss of information about long-wavelength, line-of-sight (LOS) modes in the 21\,cm data due to foregrounds, we propose a simple approach to obtain a non-vanishing signal by cross-correlating the filtered-then-squared 21\,cm brightness temperature fluctuations with the NIRB. This simple trick overcomes the mismatched Fourier space coverage of the two signals through the mode coupling induced by higher order statistics (see \citealt{Moodley2023} for the application of a similar idea to the cross-correlation of CMB lensing and post-reionization 21\,cm IM). While other methods of reconstructing the lost 21\,cm modes exist \cite[e.g.,][]{Zhu2018,Kennedy2024,Sabti2024}, we note that our approach provides a straightforward solution to directly probe the reionization timeline of interest by reinstating the complementarity of 21\,cm and NIRB signals. The 21\,cm$^2$--NIRB cross-correlation considered here relates to the 21\,cm--21\,cm--NIRB cross-bispectrum. The simpler two-point statistic encodes some of the higher-point correlation information yet involves only an angular cross-power spectrum analysis. It is therefore easier to measure, and potentially simpler to interpret, than a full cross-bispectrum analysis. Similar ideas are employed in the context of the kinetic Sunyaev-Zel'dovich (kSZ) signal, where kSZ$^2$-galaxy cross-power spectrum analyses have been proven successful \citep{Dore2004,Hill2016,LaPlante2022}. Second, we apply our method to forecast the 21\,cm$^2$--NIRB cross-correlation in multiple reionization scenarios, showcase its sensitivity to the reionization timeline, and estimate its detectability at different reionization stages by a synergy between SKA-Low and the SPHEREx deep fields. We also combine mock signals from simulations and analytic arguments based perturbative calculations to investigate the physical mechanism that connects the evolution of this cross-correlation with the reionization timeline. 

The remainder of this paper is organized as follows. In Section~\ref{sec:models}, we describe the key modeling steps and theoretical perspectives pertaining to the simulations and our analysis of the 21\,cm--NIRB cross-correlation. In Section~\ref{sec:results}, we show our main results on the cross-correlation signal, together with its detectability, in different scenarios and stages of cosmic reionization. We discuss some noteworthy implications and caveats of the analysis presented and outline a few directions for future studies in Section~\ref{sec:discussion}, before concluding in Section~\ref{sec:conclusions}. A flat, $\Lambda$CDM cosmology consistent with measurements by \citet{Planck_2016} is adopted. 

\section{Models and Theoretical Underpinnings} \label{sec:models}

In what follows, we specify how the 21\,cm and NIRB signals are self-consistently modeled in our LIMFAST simulations (Section~\ref{sec:models:limfast}), the reasons that a direct cross-correlation between 21\,cm and NIRB fluctuations would vanish in practice due to 21\,cm foreground  cleaning (Section~\ref{sec:models:direct_cross}), a simple trick to remedy this issue by squaring the foreground-filtered 21\,cm signal before cross-correlating with the NIRB (Section~\ref{sec:models:squared_cross}), and we forecast the detectability of this 21\,cm$^2$--NIRB cross-correlation (Section~\ref{sec:models:detectability}). 

\subsection{Modeling the 21\,cm Line and the NIRB Self-consistently in LIMFAST} \label{sec:models:limfast} 

\subsubsection{The 21\,cm signal of neutral hydrogen} \label{sec:models:limfast:21cm}

The modeling of the 21\,cm signal in this paper follows exactly the implementation of LIMFAST introduced by \citet{LIMFAST2023a} and \citet{LIMFAST2023b}. It builds on a combination of the excursion set formalism and perturbation theory taken from 21cmFAST \citep{Mesinger2011,Park2019} and implements a few updates on the source modeling. For reference, we provide here a brief summary of the key differences between LIMFAST and the widely used 21cmFAST code for semi-numerically simulating the 21\,cm signal. Interested readers are referred to these papers for further details. 

In addition to the introduction of multiple emission line signals tracing star formation, there are \textit{two} notable distinctions between LIMFAST and 21cmFAST that directly affect the simulated 21\,cm signal. First, with the implementation of a physically motivated, quasi-equilibrium model for feedback-regulated star formation in high-$z$ galaxies \citep{Furlanetto2017,Furlanetto2021}, LIMFAST describes the star formation rate (SFR) of galaxies at a given halo mass using physically grounded recipes for the coupling of stellar feedback. This is in contrast to the default phenomenological approach in 21cmFAST, which is based on the integrated star formation efficiency (SFE) and a star formation timescale equal to a constant fraction of the Hubble time $1/H(z)$. Even though the observed high-$z$ UV luminosity functions and densities are roughly matched in both LIMFAST and 21cmFAST to calibrate the star formation prescription, the different ways of parameterizing the SFR can actually lead to significant difference in the mass and redshift dependence \citep{LIMFAST2023a}. The additional physics not only links prescriptions of the SFE and SFR to specific mechanisms of stellar feedback (e.g., momentum- vs energy-regulated), but also enables a simple description of the gas reservoir of galaxies and thus emission line diagnostics of the multi-phase interstellar medium \citep{LIMFAST2023b}. Meanwhile, since halo emissivities responsible for ionization, X-ray heating, and Ly$\alpha$ coupling are all scaled from the SFR in this framework, LIMFAST gives physically grounded predictions of these radiation fields that shape the 21\,cm signal at different epochs. 

Second, thanks to the physically-motivated galaxy formation model included, LIMFAST also accounts for the spectral evolution of galaxies as they become chemically enriched over time --- an effect that 21cmFAST neglects. As demonstrated in \citet{LIMFAST2023a}, despite a modest effect, invoking metallicity-dependent galaxy spectral energy distribution (SED) modeling can alter both the timing and the amplitude of the 21\,cm signal. Overall, compared to previous literature, 21\,cm signals in this work are derived from semi-numerical simulations of the EoR, similar to other studies based on 21cmFAST. Our semi-numerical approach is thus complementary to the methods used in previous studies of the 21\,cm--NIRB cross-correlation, including N-body and radiative transfer simulations \citep{Fernandez2014} and the halo model \citep{Mao2014}. 

Finally, we note that the effects of molecular-cooling minihalos and Population~III stars that are available in the latest version of 21cmFAST \citep{Qin2020,Munoz2022} are neglected in this work. We do not expect their presence to significantly impact our analysis since we mainly focus on the reionization era ($z\lesssim10$) by which the radiation backgrounds associated with massive Population~III stars starts to be subdominant \citep{Mirocha2018,Mebane2020}. We also ignore redshift-space distortion effects from peculiar motions of galaxies and the IGM in this work.

\subsubsection{The NIRB} \label{sec:models:limfast:nirb}

To model the NIRB, we adopt a method similar to \citet{Mirocha2022}, but continue working with the conditional halo mass function specified by the density field rather than with individually identified halos in LIMFAST. We first use LIMFAST to generate coeval boxes of continuum emission in the observing band $\nu_1 < \nu < \nu_2$ at different redshifts by coupling the luminosity $L_{\nu^\prime}$ to the conditional halo mass function in each cell, where $\nu^\prime = \nu(1+z)$ is the rest-frame emission frequency corresponding to the observed frequency $\nu$ and $L_{\nu^\prime}$ is a function of halo mass, redshift, and metallicity. We forward model $L_{\nu^\prime}$ by creating galaxy SEDs with the stellar population synthesis code
BPASS \citep{Eldridge2017}, assuming a composite stellar population of single stars with a constant star formation history\footnote{The dominant presence of galaxies with bursty star formation histories at $z\gtrsim6$ recently discovered by JWST may impact the NIRB by introducing strong stochasticity in the source luminosity and increasing the strength of the nebular emission \citep{Sun2023NIRB,Katz2024}. While such effects associated with non-constant star formation histories are not captured in the density field-based LIMFAST simulations used here, we plan to look into them using halo-based LIMFAST simulations in future studies.}
and then processing them with the photoionization code \textsc{cloudy} \citep{Ferland2017}, such that the resulting net SED self-consistently captures the contributions from stellar emission, nebular lines (e.g., Lyman and Balmer series lines), and nebular continuum (including two-photon, free-free, and free-bound radiations; \citealt{Sun2021NIRB}). The impact of dust attenuation on galaxy SEDs is neglected, since at $z \gtrsim 6$ it has been observed to remain relatively modest even in massive galaxies \citep{Bowler2024} and the obscured fraction of total cosmic star formation is low \citep{Zavala2021}. Intensities from coeval boxes at different redshifts are then interpolated and stacked into a light cone, before being summed over along the line of sight (with effects like cosmological redshift and IGM absorption properly accounted for) to get the NIRB maps. 

Following \citet{Sun2021NIRB}, we can express the band-averaged NIRB mean intensity contributed by sources above redshift $z_0$ as
\begin{equation}
    I_{\nu} = \bar{I}_{\nu_1 \nu_2} = \frac{1}{\Delta \nu} \int_{z_0} d z \frac{d I}{d z} = \frac{c}{\Delta \nu} \int_{z_0} d z \frac{\int_{\nu'_1}^{\nu'_2} d \nu' \int d M \hat{L}_{\nu'}(M,z,Z) \frac{dn}{dM}}{4\pi H(z)(1+z)^2} = \frac{c}{\Delta \nu} \int_{z_0} d z \frac{\dot{\rho}_*(z) \int_{\nu'_1}^{\nu'_2} d \nu' l_{\nu'}(z, Z) e^{-\tau_{\nu'}(z)}}{4\pi H(z)(1+z)^2},
\end{equation}
where $c$ is the speed of light, $\Delta \nu = \nu_2 - \nu_1$ is the observing bandpass, $H(z)$ is the Hubble parameter, and the numerator of the integrand is evaluated for the rest-frame frequency $\nu'$ throughout. The luminosity attenuated by the intergalactic neutral hydrogen at $z>z_0$ is $\hat{L}_{\nu'}(M,z,Z)=L_{\nu'}(M,z,Z) e^{-\tau_{\nu'}(z)}=l_{\nu'}(z, Z) \dot{M}_*(M,z)e^{-\tau_{\nu'}(z)}$, where $\tau_{\nu'}$ is the net optical depth at frequency $\nu'$ due to the neutral IGM. For simplicity, we assume $\tau_{\nu'} = \infty$ (complete absorption) if $\nu' > \nu_\mathrm{Ly\alpha}$ and 0 (no absorption) otherwise. More accurate treatment of the IGM transmission can be done with analytic models motivated by observations \cite[e.g.,][]{Inoue2014}, but should not impact the results of our analysis in any significant way because wavelengths longer than Ly$\alpha$ are little affected by IGM transmission effects.

\subsection{A Vanishing 21\,cm--NIRB Cross-correlation} \label{sec:models:direct_cross}

Considering spatial fluctuations in the transverse and LOS directions, we can write down the expression of a direct cross-correlation between the 3D, foreground-filtered 21\,cm brightness temperature field $\mathcal{T}$ and the contribution to the two-dimensional (2D) NIRB intensity field from the same redshift $\mathcal{I} = d I_{\nu} / d z$ as
\begin{equation}
P_{\mathcal{T},\mathcal{I}}(k_{\parallel}, \boldsymbol{k}_{\perp}) = \langle \tilde{\mathcal{T}}(k_{\parallel}, \boldsymbol{k}_{\perp}) \tilde{\mathcal{I}}(0, \boldsymbol{k}^\prime_{\perp}) \rangle = (2\pi)^3 P_{T,\mathcal{I}}(0, \boldsymbol{k}_{\perp}) \delta_{D}(k_{\parallel}) \delta_{D}(\boldsymbol{k}_{\perp} + \boldsymbol{k}^\prime_{\perp}) W_\mathrm{fg}(k_\parallel),
\label{eq:TI_vanish}
\end{equation}
where $\tilde{\mathcal{I}}(0, \boldsymbol{k}_{\perp})$ denotes the Fourier transform of the NIRB intensity with purely the transverse component, whereas $\tilde{\mathcal{T}}(k_{\parallel}, \boldsymbol{k}_{\perp}) = \tilde{T}(k_{\parallel}, \boldsymbol{k}_{\perp}) W_\mathrm{fg}(k_\parallel)$ denotes the Fourier transform of the 21\,cm signal filtered by a sharp-$k$, high-pass filter $W_\mathrm{fg}$ to reject long-wavelength LOS modes contaminated by the foregrounds that have slowly varying frequency structures. Specifically, we can express $W_\mathrm{fg}$ as
\begin{equation}
W_\mathrm{fg}(k_\parallel) = 
    \begin{cases}
     1~, & k_\parallel > k_\mathrm{\parallel,fg} \\[1\jot]
     0~, & k_\parallel \leq k_\mathrm{\parallel,fg}
    \end{cases}
\label{eq:sharpk}
\end{equation}
in the $k_\parallel$ direction to remove the lowest two or three line-of-sight modes of the 21\,cm power spectrum with frequencies below $k_{\parallel,\mathrm{fg}}$, which are highly dominated by foregrounds. 

Because $W_\mathrm{fg}(k_{\parallel} \rightarrow 0) = 0$ by definition, directly cross-correlating $\mathcal{T}$ and $\mathcal{I}$ for transverse modes with $k_{\parallel} = 0$---the only modes probed by the NIRB---yields a vanishing result due to the mismatched Fourier space coverage. Therefore, the foreground cleaning required for the 21\,cm signal prohibits a direct 21\,cm--NIRB cross-correlation from being detected. Some alternative means to cross-correlate the two fields is needed to extract the complementary information about the EoR encoded in them. 

\subsection{A Non-vanishing Cross-correlation Between 21\,cm$^2$ and NIRB} \label{sec:models:squared_cross}

As recently discussed in \citet{Moodley2023}, turning to higher order statistics like the (cross-)bispectrum can prevent the cross-correlation signal from vanishing when cross-correlating the foreground-filtered 21\,cm IM data with projected (2D) large-scale structure tracers like the CMB lensing signal. Here we explore and demonstrate a similar idea in the context of the 21\,cm and NIRB signals during the EoR. Rather than working with the full cross-bispectrum statistic, we focus on a simple alternative, which uses the \textit{squared} 21\,cm field. The corresponding cross-correlation remains a two-point statistic that can be more easily measured and interpreted, but the 21\,cm$^2$--NIRB cross-power spectrum nevertheless exploits the non-vanishing higher-order correlations between the unfiltered high-$k_\parallel$ 21\,cm power spectrum  and NIRB fluctuations. As we show below, this new cross-power spectrum may formally be expressed in terms of integrals over the cross-bispectrum (whose connection with the physics of reionization can be understood perturbatively; see Section~\ref{sec:discussion:sign_change})

Considering the squared, foreground-filtered 21\,cm signal, $\mathcal{T}^2$, we can show how its cross-correlation with the NIRB in this case corresponds to a projected cross-bispectrum that does not vanish for $k_{\parallel} \rightarrow 0$. The cross-bispectrum considered here quantifies correlations between the high-$k_{\parallel}$ 21\,cm power spectrum and the $k_{\parallel} \rightarrow 0$ fluctuations in the NIRB. Essentially, it quantifies whether the small-scale 21\,cm fluctuation power spectrum is enhanced or reduced within NIRB bright spots. With the same foreground filter $W_\mathrm{fg}$ as previously defined, we define $\mathcal{A} \equiv \mathcal{T}^2$ to be the square of the filtered 21\,cm signal, such that
\begin{equation}
\tilde{\mathcal{A}}(\boldsymbol{k}) = \tilde{\mathcal{T}^2}(\boldsymbol{k}) = \int \frac{d^3 q}{(2\pi)^3} \tilde{\mathcal{T}}(\boldsymbol{k}-\boldsymbol{q}) \tilde{\mathcal{T}}(\boldsymbol{q}) = \int \frac{d q_{\parallel}}{2\pi} \frac{d^2 q_{\perp}}{(2\pi)^2} W_\mathrm{fg}(q_{\parallel}) W_\mathrm{fg}(k_{\parallel} - q_{\parallel}) \tilde{T}(q_{\parallel}, \boldsymbol{q}_{\perp}) \tilde{T}(k_{\parallel} - q_{\parallel}, \boldsymbol{k}_{\perp}-\boldsymbol{q}_{\perp}),
\end{equation}
which implies
\begin{equation}
\tilde{\mathcal{A}}(k_{\parallel}=0, \boldsymbol{k}_{\perp}) = \int \frac{d q_{\parallel}}{2\pi} \frac{d^2 q_{\perp}}{(2\pi)^2} |W_\mathrm{fg}(q_\mathrm{\parallel})|^2 \tilde{T}(q_{\parallel}, \boldsymbol{q}_{\perp}) \tilde{T}(- q_{\parallel}, \boldsymbol{k}_{\perp}-\boldsymbol{q}_{\perp}).
\end{equation}
Note that the integral also picks up from the non-vanishing part of $W_\mathrm{fg}$, which is crucial for the cross-correlation with the NIRB. Given that
\begin{equation}
\langle \tilde{\mathcal{A}}(0, \boldsymbol{k}_{\perp}) \tilde{\mathcal{I}}(0,\boldsymbol{k}^\prime_{\perp}) \rangle = (2\pi)^3 \delta_{D}(\boldsymbol{k}_{\perp} + \boldsymbol{k}^\prime_{\perp}) P_{\mathcal{A}, \mathcal{I}}(0, \boldsymbol{k}_{\perp})
\end{equation}
and
\begin{equation}
\langle \tilde{\mathcal{A}}(0, \boldsymbol{k}_{\perp}) \tilde{\mathcal{I}}(0,\boldsymbol{k}^\prime_{\perp}) \rangle = \int \frac{d q_{\parallel}}{2\pi} \frac{d q_{\perp}}{(2\pi)^2} |W_\mathrm{fg}(q_\mathrm{\parallel})|^2 \langle \tilde{T}(-q_{\parallel}, \boldsymbol{k}_{\perp} - \boldsymbol{q}_{\perp}) \tilde{T}(q_{\parallel}, \boldsymbol{q}_{\perp}) \tilde{\mathcal{I}}(0, \boldsymbol{k}^\prime_{\perp}) \rangle,
\end{equation}
with
\begin{equation}
\langle \tilde{T}(-q_{\parallel}, \boldsymbol{k}_{\perp} - \boldsymbol{q}_{\perp}) \tilde{T}(q_{\parallel}, \boldsymbol{q}_{\perp}) \tilde{\mathcal{I}}(0, \boldsymbol{k}^\prime_{\perp}) \rangle = (2\pi)^2 \delta_{D}(\boldsymbol{k}_{\perp} + \boldsymbol{k}^\prime_{\perp}) B_{T, T, \mathcal{I}}\left( -q_{\parallel} \hat{\boldsymbol{z}}+\boldsymbol{k}_{\perp} - \boldsymbol{q}_{\perp},\ q_{\parallel} \hat{\boldsymbol{z}}+\boldsymbol{q}_{\perp},\ \boldsymbol{k}^\prime_{\perp} \right),
\end{equation}
where $B_{T,T,\mathcal{I}}$ denotes the cross-bispectrum between two 21\,cm fields and the NIRB fluctuations, we can show that the cross-power spectrum of $\mathcal{A}$ and $\mathcal{I}$ is essentially a projected cross-bispectrum, namely
\begin{equation}
P_{\mathcal{A}, \mathcal{I}}(0, \boldsymbol{k}_{\perp}) = \int \frac{d q_{\parallel}}{2\pi} \frac{d q_{\perp}}{(2\pi)^2} |W_\mathrm{fg}(q_\mathrm{\parallel})|^2 B_{T, T, \mathcal{I}}\left( -q_{\parallel} \hat{\boldsymbol{z}}+\boldsymbol{k}_{\perp} - \boldsymbol{q}_{\perp},\ q_{\parallel} \hat{\boldsymbol{z}}+\boldsymbol{q}_{\perp},\ \boldsymbol{k}^\prime_{\perp} \right).
\label{eq:bs_proj}
\end{equation}
Through the higher-point correlator $B_{T,T,\mathcal{I}}$, the projection induces mode mixing that couples high-frequency/small-scale $k_{\parallel}$ modes of the 21\,cm fluctuations that survive the foreground filtering to the NIRB fluctuations measured purely in transverse directions. 

By expanding the power spectrum $P(k_{\parallel},\boldsymbol{k}_{\perp})$ about $k_{\parallel}=0$ in the flat sky limit, we can show that under the Limber approximation the angular cross-power spectrum $C_{\mathcal{A},I}(\ell)$, or the 2D cross-power spectrum $P^\mathrm{2D}_{\mathcal{A},I}$, is given by the projection
\begin{equation}
\ell(\ell+1)C_{\mathcal{A},I}(\ell) \approx k^2_{\perp} P^\mathrm{2D}_{\mathcal{A},I}(k_{\perp}) = k^2_{\perp} c \int \frac{F_{\mathcal{A}}(z)  F_{\mathcal{I}}(z) P_{\mathcal{A},\mathcal{I}}\left[ k_{\parallel}=0, k_{\perp}=\ell/r(z) \right]}{H(z) r^2(z)} d z,
\end{equation}
where $r(z)$ is the comoving radial distance and the multipole moment $\ell \approx k_{\perp} r(z)$ for $\ell \gg 1$. The projection kernels $F_\mathcal{A} = 1/(2\Delta z)$ for 21\,cm signals in a redshift range $z \pm \Delta z$ and $F_{\mathcal{I}} = 1$ for the NIRB. Similarly, for the squared 21\,cm field, the angular auto-power spectrum is
\begin{equation}
\ell(\ell+1)C_{\mathcal{A}}(\ell) \approx k^2_{\perp} P^\mathrm{2D}_{\mathcal{A}}(k_{\perp}) = k^2_{\perp} c \int \frac{F^2_{\mathcal{A}}(z) P_{\mathcal{A}}\left[ k_{\parallel}=0, k_{\perp}=\ell/r(z) \right]}{H(z) r^2(z)} d z. 
\end{equation}

\subsection{Detectability of the 21\,cm$^2$--NIRB Cross-power Spectrum} \label{sec:models:detectability}

The uncertainty of the angular cross-power spectrum has contributions from the sample variance and measurement uncertainties of the respective signals. Under the flat-sky approximation, we express the angular cross-power spectrum uncertainty as
\begin{equation}
\Delta C_{\mathcal{A},I}(\ell) = \left\{\frac{C^2_{\mathcal{A},I}(\ell)+\left[C_{\mathcal{A}}(\ell) + C_{N,\mathcal{A}}(\ell)\right] \left[C_{I}(\ell) + C_{N,I}(\ell)\right]}{(2\ell+1)f_\mathrm{sky}\Delta \ell} \right\}^{1/2}, 
\label{eq:cps_err}
\end{equation}
where $C_{I}(\ell)$ and $C_{N,I}(\ell)$ are the angular power spectrum of the NIRB intensity and its uncertainty due to instrument noise. The uncertainty of the $\mathcal{A}$ field, $C_{N,\mathcal{A}}(\ell)$, associated with 21\,cm instrument noise can be expressed as
\begin{equation}
\ell(\ell+1)C_{N,\mathcal{A}}(\ell) \approx k^2_{\perp} P^\mathrm{2D}_{N,\mathcal{A}}(k_{\perp}) = k^2_{\perp} c \int \frac{F^2_{\mathcal{A}}(z) P_{N,\mathcal{A}}\left[ k_{\parallel}=0, k_{\perp}=\ell/r(z) \right]}{H(z) r^2(z)} d z.
\end{equation}
In the idealized case where uncertainties of both $\mathcal{A}$ and $I$ fields are sample variance-dominated and the cross-correlation coefficient $r_{\times} = C_{\mathcal{A},I}/\sqrt{C_{\mathcal{A}} C_I}$ is small, we can approximate the signal-to-noise ratio (S/N) of the cross power spectrum, $C_{\mathcal{A},I}$, as
\begin{equation}
\left( \mathrm{\frac{S}{N}} \right)^2_{\times} \approx 2\ell^2 f_\mathrm{sky} \Delta \ln \ell\,r_{\times}^2, 
\end{equation}
which amounts to roughly $20$ at $\ell = 1000$ for $r_{\times} = 0.3$, $f_\mathrm{sky} = 0.005$, and $\Delta \ln \ell = 0.5$ (see Section~\ref{sec:results:detectability} for detailed discussions of the detectability of $C_{\mathcal{A},I}$ using the SKA and SPHEREx). In reality, however, both instrumental noise and residual contamination from foregrounds contribute to the auto-correlation variance terms and make the detection of the cross-correlation signal more challenging. In Section~\ref{sec:results:detectability}, we will show the detectability of $C_{\mathcal{A},I}$ estimated based on the signal predicted by LIMFAST and the instrument noises of a 1000-hour SKA-Low survey in overlap with the 200\,deg$^2$ SPHEREx deep fields. 

\begin{figure*}[!t]
 \centering
 \includegraphics[width=0.48\textwidth]{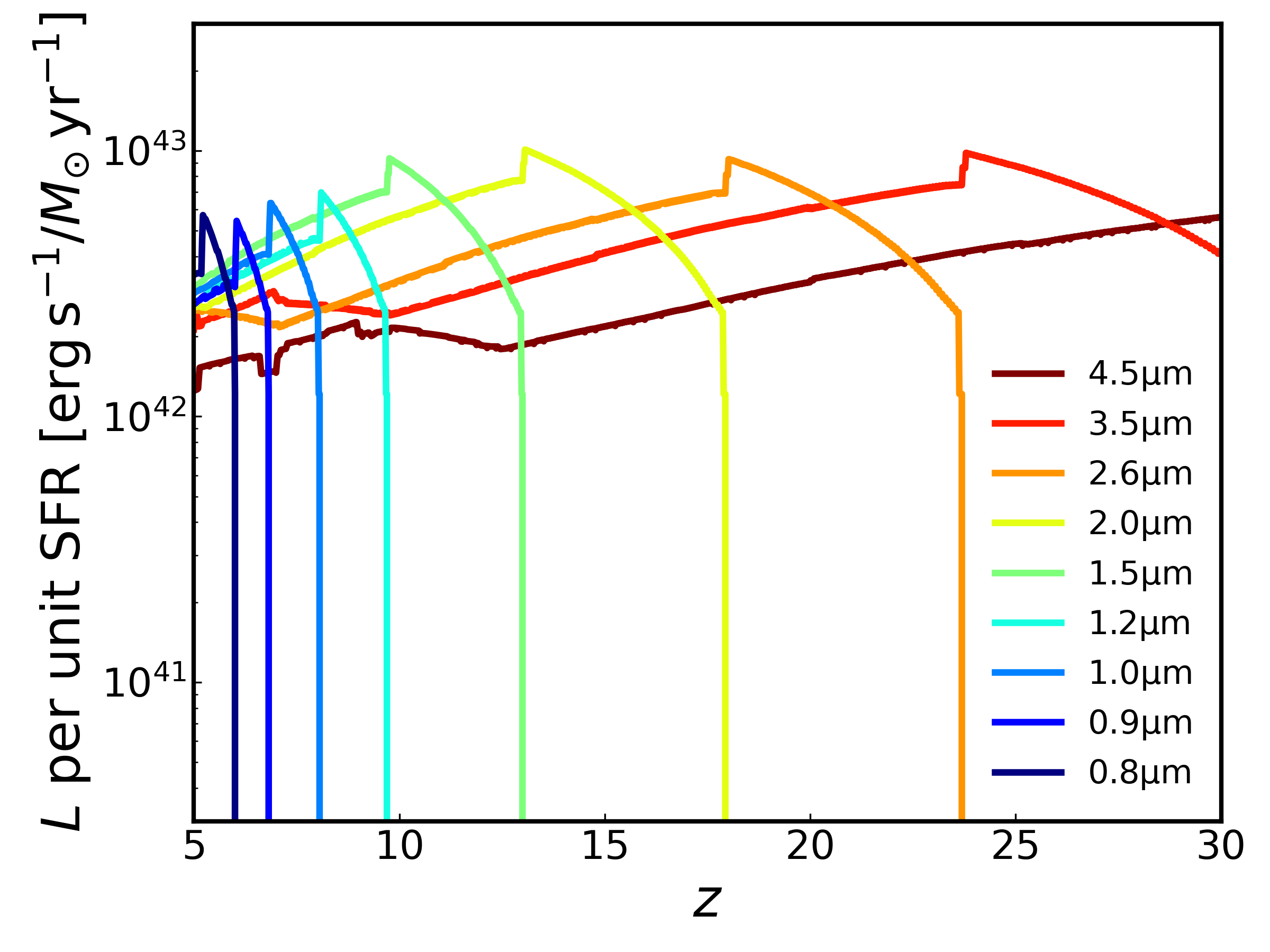}
 \includegraphics[width=0.48\textwidth]{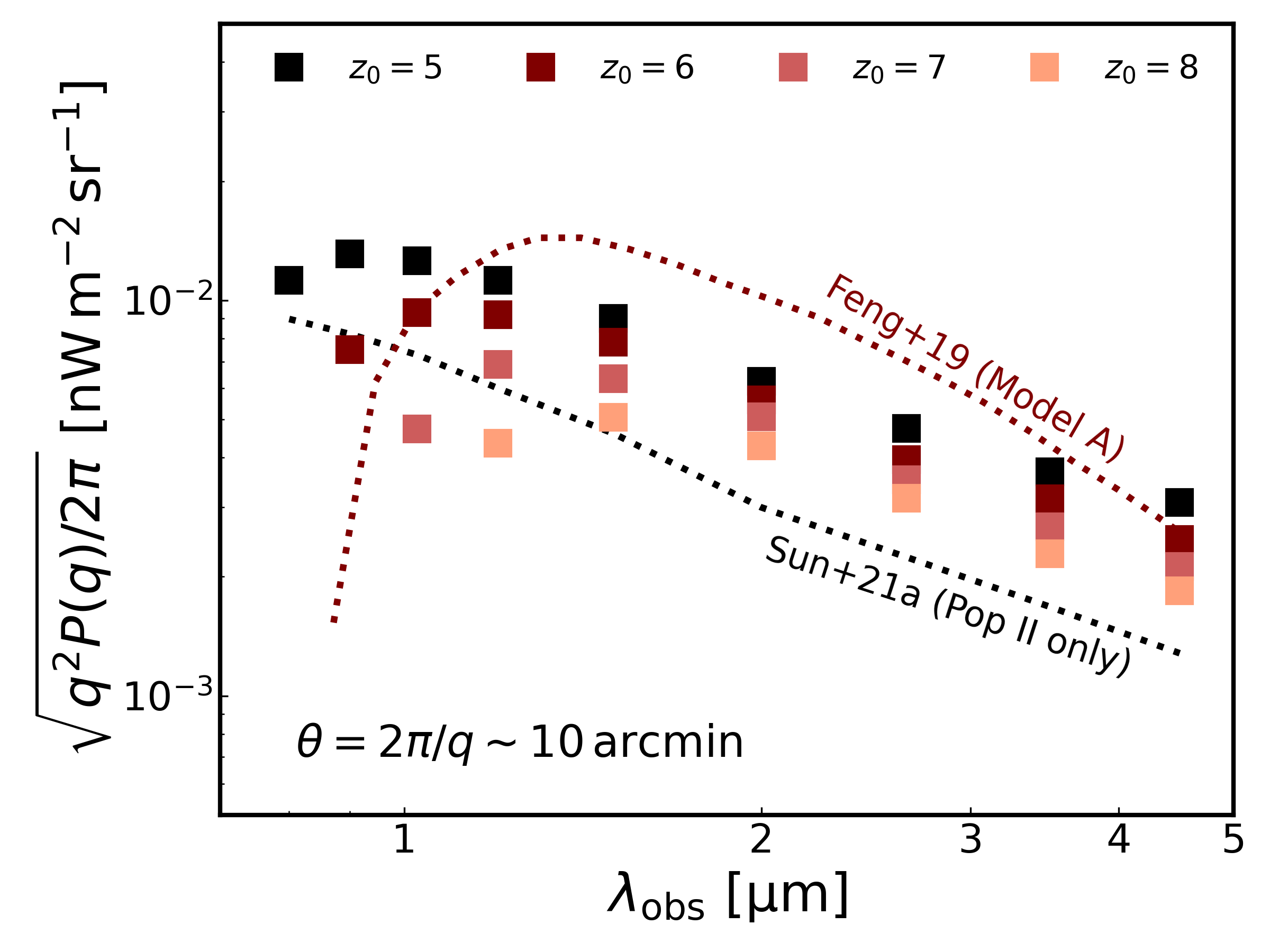}
 \caption{Left: band-integrated NIRB luminosity per unit SFR predicted by \textsc{cloudy}-processed BPASS galaxy SEDs in 9 example wavelength bands as a function of redshift. The sharp cutoff at the high redshift end of each band (e.g., $z\sim13$ for the 1.5\,$\mu$m band) indicates where the Ly$\alpha$ break redshifts out of the bandpass. The overall shape of the curves displayed depends on the wavelength and bandwidth, with the contribution to longer wavelength bands extending to higher redshift. Right: the NIRB angular fluctuations on approximately 10$\arcmin$ scales measured from mock NIRB maps simulated by LIMFAST in the 9 example wavelength bands. Different colors indicate different minimum redshifts for calculating the NIRB, with the Ly$\alpha$ break causing a dropout effect in short-wavelength bands starting $z_0 \gtrsim 6$. For comparison, the dotted curves show model predictions for $z_0 = 5$ from \citet{Sun2021NIRB} and $z_0 = 6$ from \citet{Feng2019}.}
 \label{fig:NIRB}
\end{figure*}

\section{Results} \label{sec:results}

\subsection{The 21\,cm and NIRB Signals During the EoR}

\begin{figure*}[!t]
 \centering
 \includegraphics[width=\textwidth]{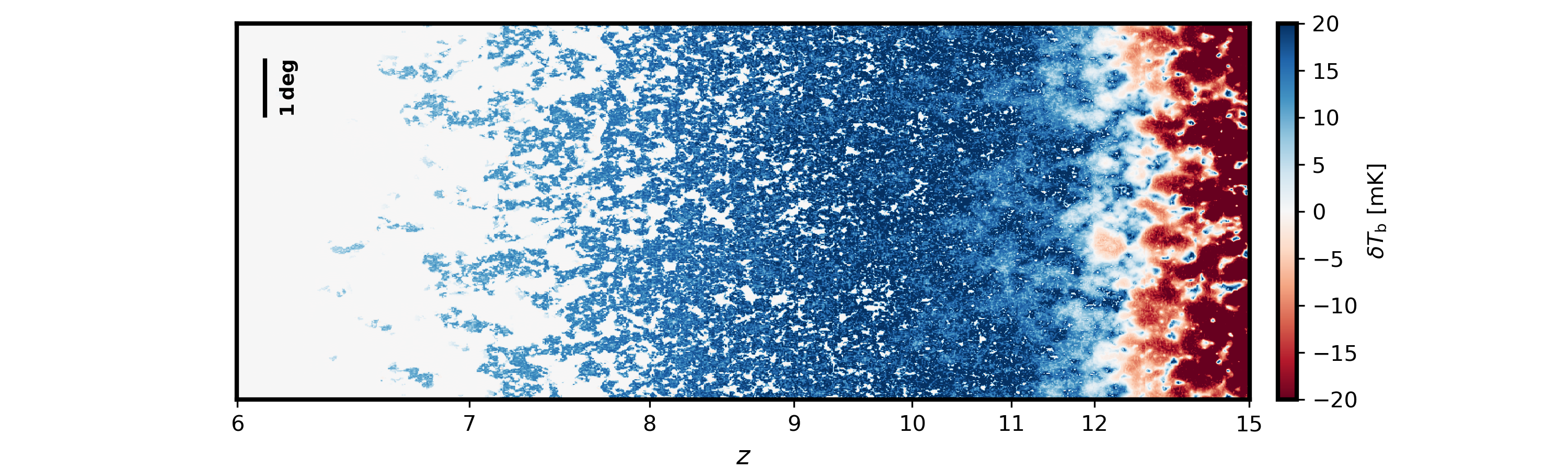}
 \includegraphics[width=\textwidth]{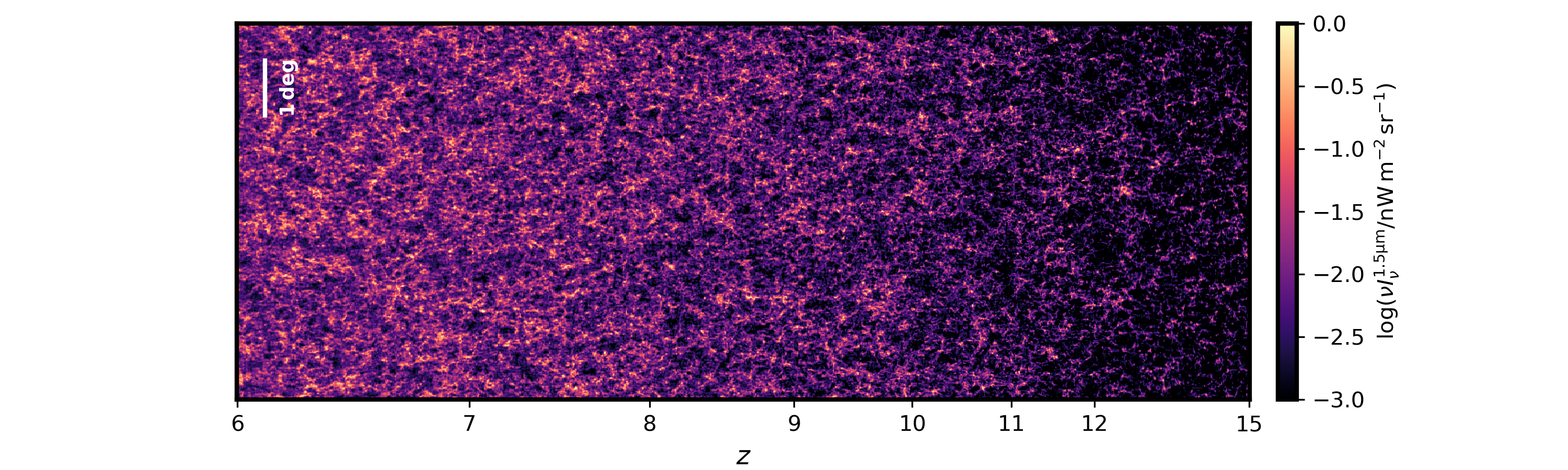}
 \caption{Mock $7\times7\ \mathrm{deg^2}$ light cones of the 21\,cm differential brightness temperature (top) and the NIRB intensity at 1.5\,$\mu$m (bottom) produced by LIMFAST assuming the fiducial, momentum-driven feedback from supernovae, model. Thanks to the large volume of the coeval boxes (1024\,cMpc$^3$), only two boxes need to be interpolated and stacked when generating a light cone from $z=15$ to $z=6$.}
 \label{fig:lightcones}
\end{figure*}

As described in Section~\ref{sec:models:limfast}, we self-consistently simulate both 21\,cm and NIRB signals with LIMFAST. The broadband nature of the NIRB signal requires contributions from different redshifts to be taken into account and combined. In the left panel of Figure~\ref{fig:NIRB}, for 9 example NIR bands covering the same wavelength range as SPHEREx, we show the band-integrated NIRB luminosity per unit SFR predicted by our \textsc{cloudy}-based galaxy SED model as a function of redshift. Due to the presence of the Ly$\alpha$ break (which gives rise to the sharp cutoff at the high-redshift end), longer wavelength bands have contributions from a wider range of redshifts. The right panel of Figure~\ref{fig:NIRB} shows the angular fluctuations of the NIRB on 10$\arcmin$ scales as a function of wavelength measured from mock NIRB maps simulated by LIMFAST. Results are displayed for a few different minimum redshifts to showcase a breakdown of the contributions as well as the effect of neutral IGM absorption that causes the spectral break at Ly$\alpha$ wavelength (apparent for $z_0 \gtrsim 6$). We also compare these predictions from LIMFAST simulations with previous results based on analytic models from \citet[][negelecting the small contribution from Population~III stars]{Sun2021NIRB} and \citet[][Model A]{Feng2019} and find decent agreement between them. The (roughly factor of 2) difference between this work and the results from \citet{Sun2021NIRB} is mainly due to the larger cosmic SFR density predicted by the fiducial, momentum-driven feedback model assumed in LIMFAST\footnote{As shown by the comparison in Figure~\ref{fig:cps_vs_xhi}, LIMFAST implements galaxy formation models with varying strengths of supernova feedback (e.g., momentum-driven and energy-driven), which impacts the reionization timeline by modulating the SFE of galaxies as a function of halo mass (see also \citealt{LIMFAST2023b}).}. 

\begin{figure*}[!ht]
 \centering
 \includegraphics[width=\textwidth]{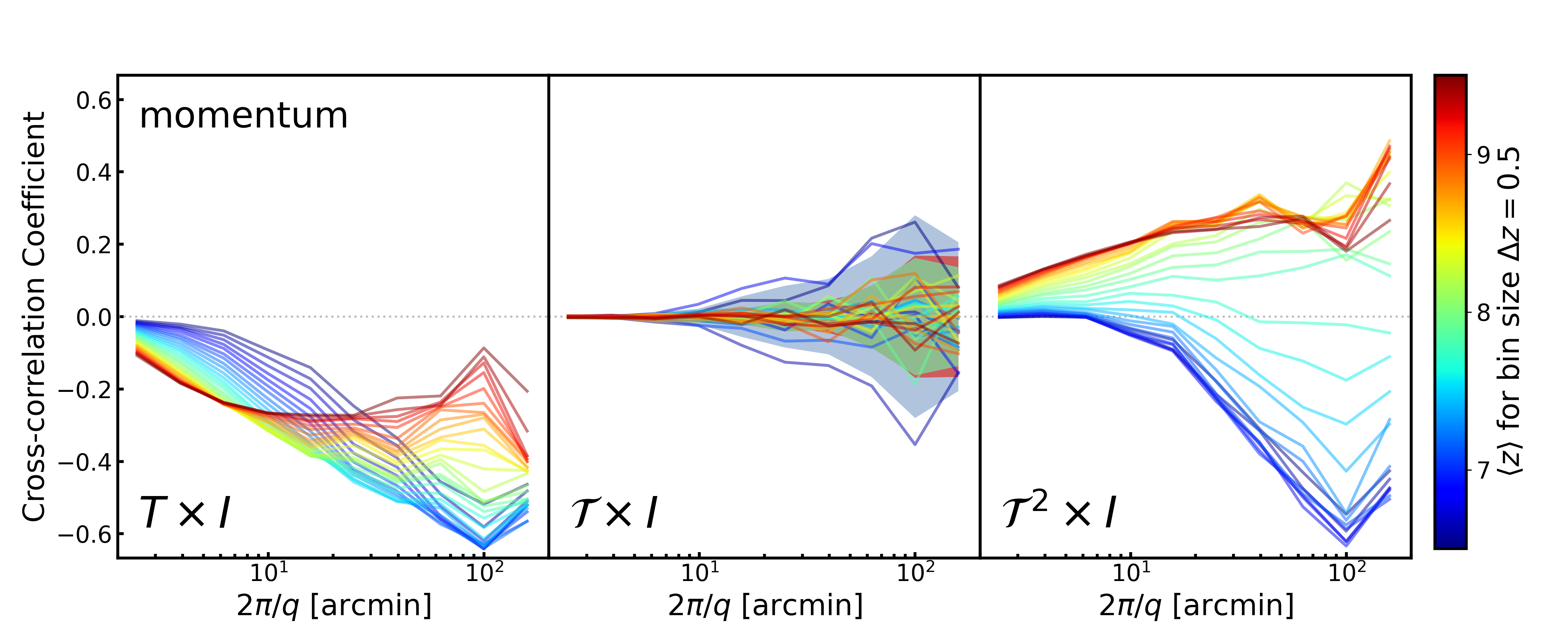}
 \includegraphics[width=\textwidth]{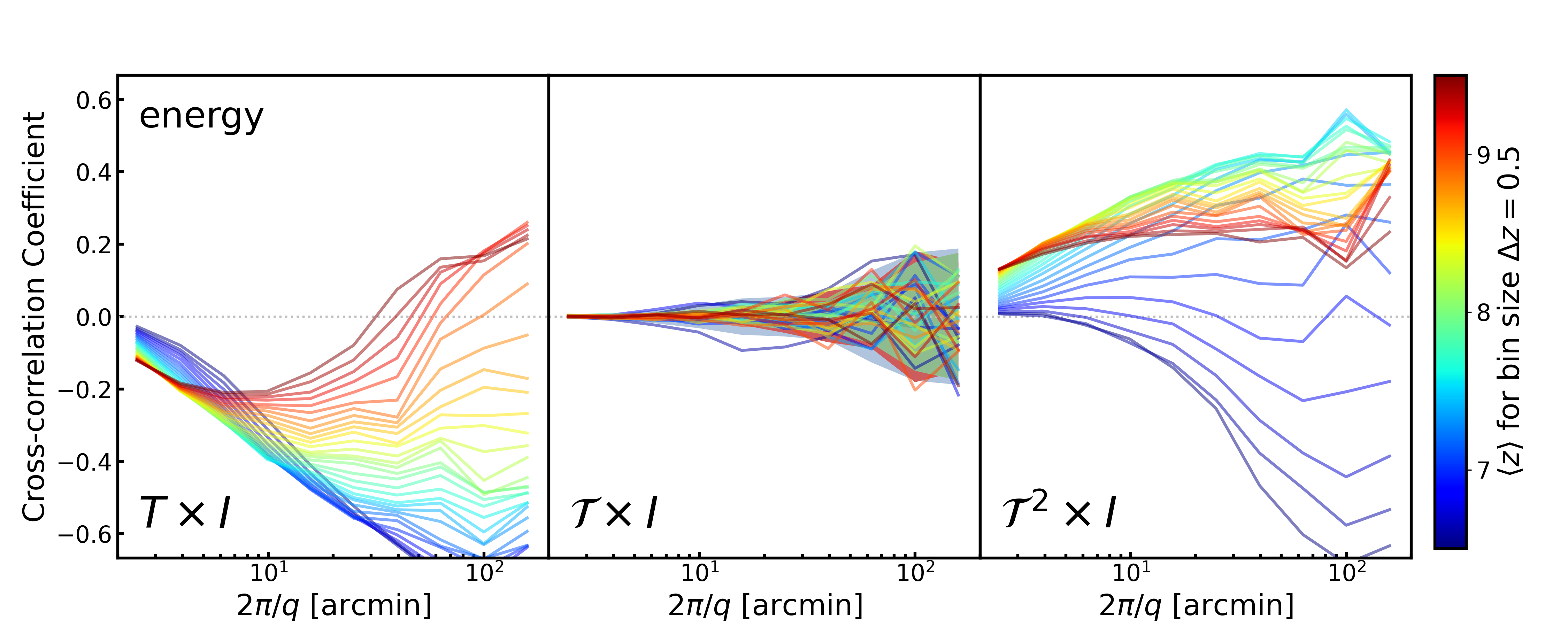}
 \caption{Top: cross-correlation coefficients between the NIRB at $1.3<\lambda<1.7\,\mu$m and the unfiltered (left), filtered (middle), and filtered-then-squared (right) 21\,cm signal as a function of the angular scale defined by the inverse wavenumber. From left to right, the 3 panels show explicitly how the vanishing cross-correlation of the NIRB and 21 cm signals caused by 21 cm foreground filtering can be reinstated by the higher order statistics. In the middle panel, we plot 3 shaded bands in the background to indicate the $\pm1\sigma$ levels of random noise in our measurements at $z\sim6.8$ (blue), 8.0 (green), and 9.3 (red), which suggest that the measured cross-correlation signals are consistent with zero on all scales at all redshifts. The results displayed are calculated from light cones covering approximately $7 \times 7$ deg$^2$ and $6<z<15$ (see Figure~\ref{fig:lightcones}) assuming momentum-driven feedback from supernovae, and are averaged over 5 random realizations. The 21\,cm signals are evaluated for redshift bins of $\langle z \rangle \pm \Delta z$ with $\Delta z = 0.5$. Bottom: same as the top panel but for more efficient, energy-driven feedback from supernovae. In both cases, the sign of the cross-correlation coefficient changes when the mean IGM neutral fraction is approximately 40\%, which is reached at a lower redshift for less efficient star formation regulated by energy-driven feedback.}
 \label{fig:ccc}
\end{figure*}

For the joint analysis of 21\,cm and NIRB signals, we run LIMFAST to generate mock realizations of them at $6 < z < 15$. In each realization, original coeval boxes of the 21\,cm and NIRB signals have a volume of 1024\,cMpc$^3$ and a spatial resolution of 2\,cMpc (i.e., each box has $512^3$ cells), and light cones are created by interpolating the coeval boxes at discretely sampled redshifts, after flipping them in random sequences to minimize repeating structures. To better understand the impact of galaxy formation physics on our results, as in \citet{LIMFAST2023b}, we consider two different modes of stellar feedback, momentum-driven vs energy-driven (see also the discussion in \citealt{Furlanetto2017}), that predict slightly different reionization scenarios in terms of both timeline and morphology. We run 5 random realizations for each choice of feedback mode in order to average out some numerical uncertainties and more robustly determine any trends. 

To visualize the relationship between 21\,cm and NIRB signals, we show in Figure~\ref{fig:lightcones} mock light cones of them over $6 < z < 15$ simulated by LIMFAST assuming momentum-driven feedback from supernovae. Since the 21\,cm and NIRB signals in general trace the neutral gas and ionizing sources respectively, their strengths appear to be negatively correlated for the most part. However, as will be shown later, the exact cross-correlation between them has more complicated and physically informative behavior. 

\subsection{The 21\,cm$^2$--NIRB Cross-correlation} \label{sec:results:cross}

In Sections~\ref{sec:models:direct_cross} and \ref{sec:models:squared_cross}, we have analytically shown that a direct cross-correlation between the foreground-filtered 21\,cm signal $\mathcal{T}$ and the NIRB signal $I$ would vanish, whereas a cross-correlation between $\mathcal{T}^2$ and $I$ would not thanks to the mode coupling between the 21\,cm and NIRB fields, which leads to a non-vanishing cross-bispectrum. Here, we demonstrate the validity of these arguments using mock data simulated by LIMFAST. Figure~\ref{fig:ccc} shows the 21\,cm--NIRB cross-correlation coefficient as a function of angular scales and redshift. For simulations generated for each of the two stellar feedback modes, 3 cases are shown: (1) the left panel shows the direct cross-correlation between the \textit{unfiltered} 21 cm signal and the NIRB; (2) the middle panel shows the direct cross-correlation between the foreground \textit{filtered} 21\,cm signal and the NIRB; and (3) the right panel shows the cross-correlation between the \textit{filtered-then-squared} 21\,cm signal and the NIRB. Each cross-correlation is measured between the 2D NIRB map and the 21\,cm fluctuations in a redshift layer $\langle z \rangle \pm \Delta z$. 

\begin{figure*}[!ht]
 \centering
 \includegraphics[width=0.325\textwidth]{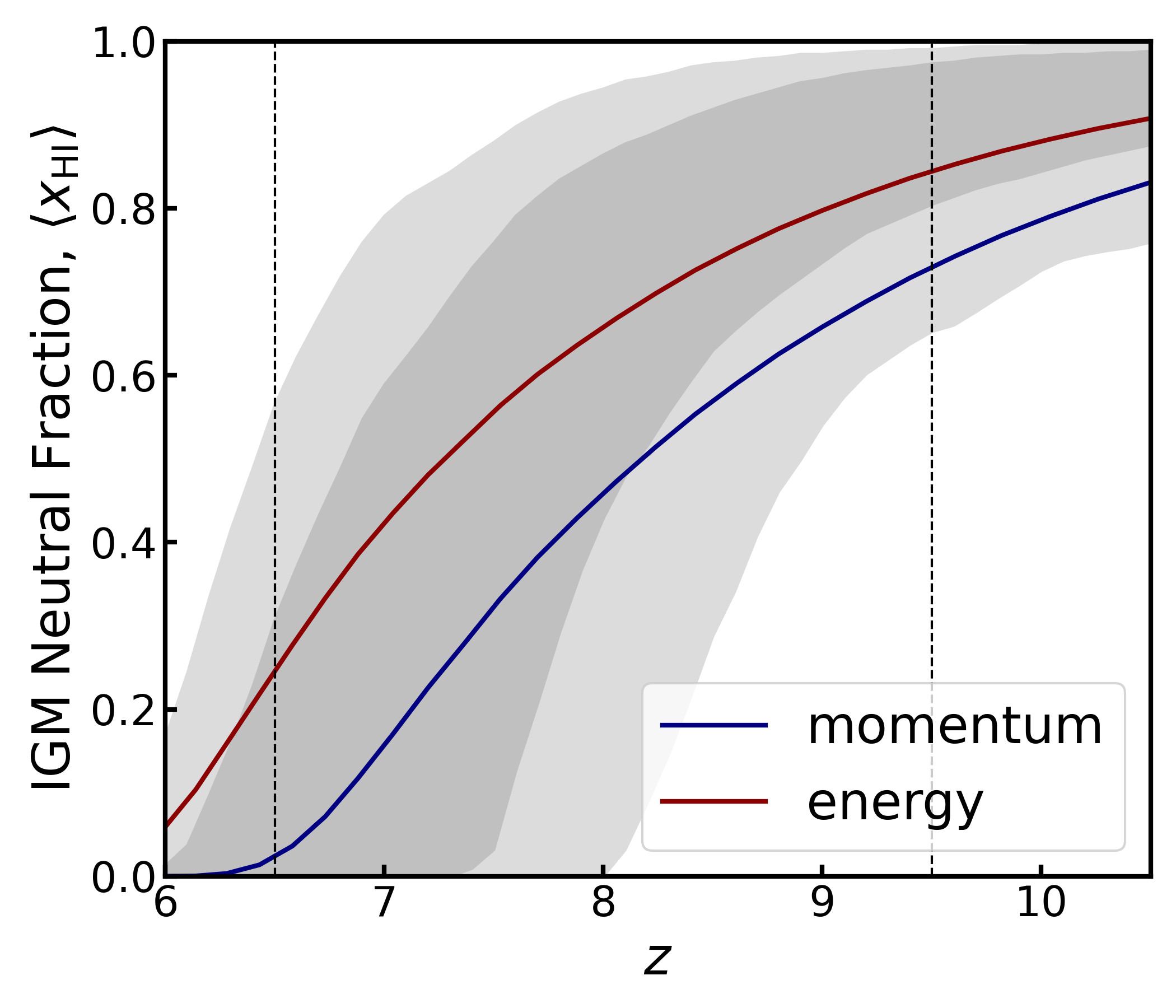}
 \includegraphics[width=0.325\textwidth]{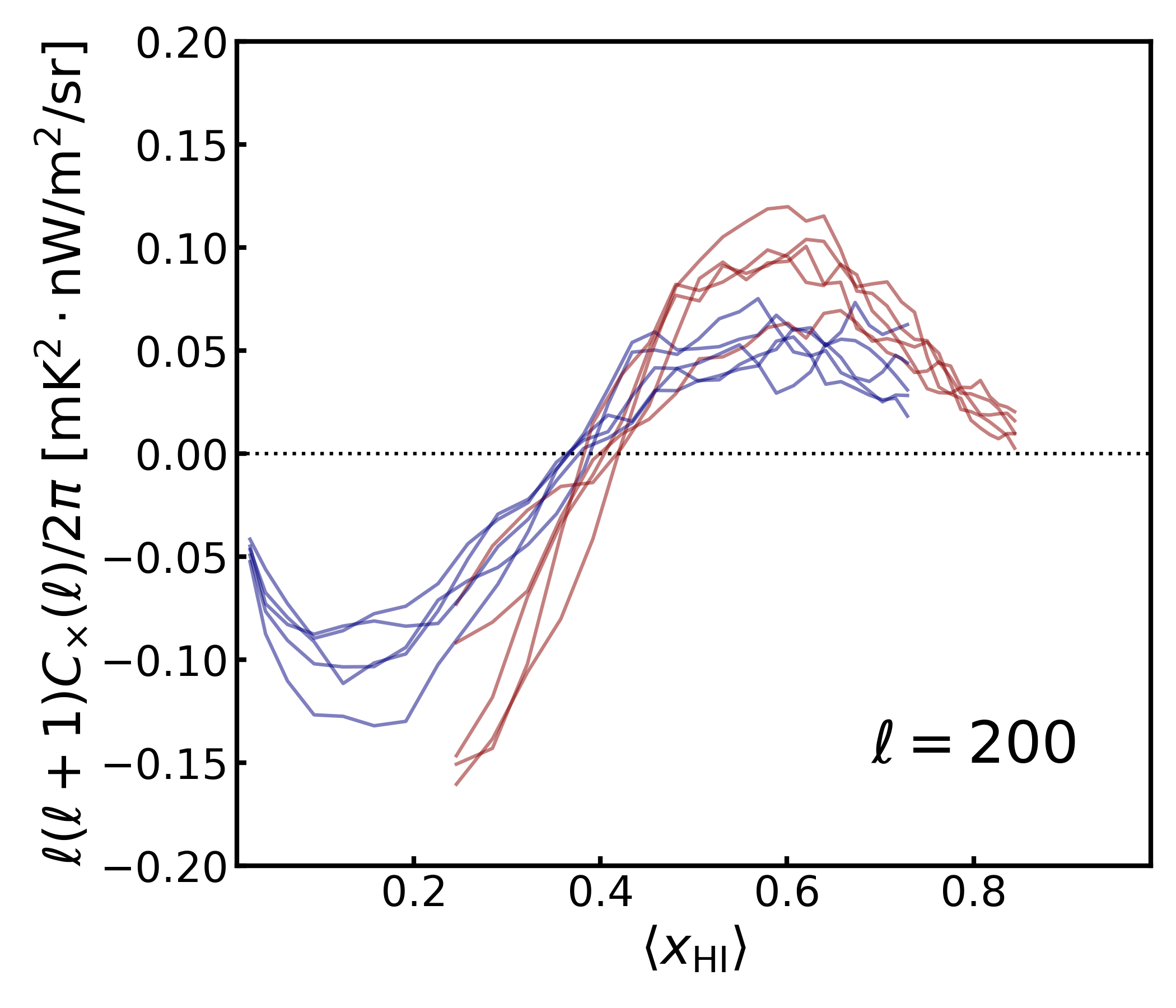}
 \includegraphics[width=0.325\textwidth]{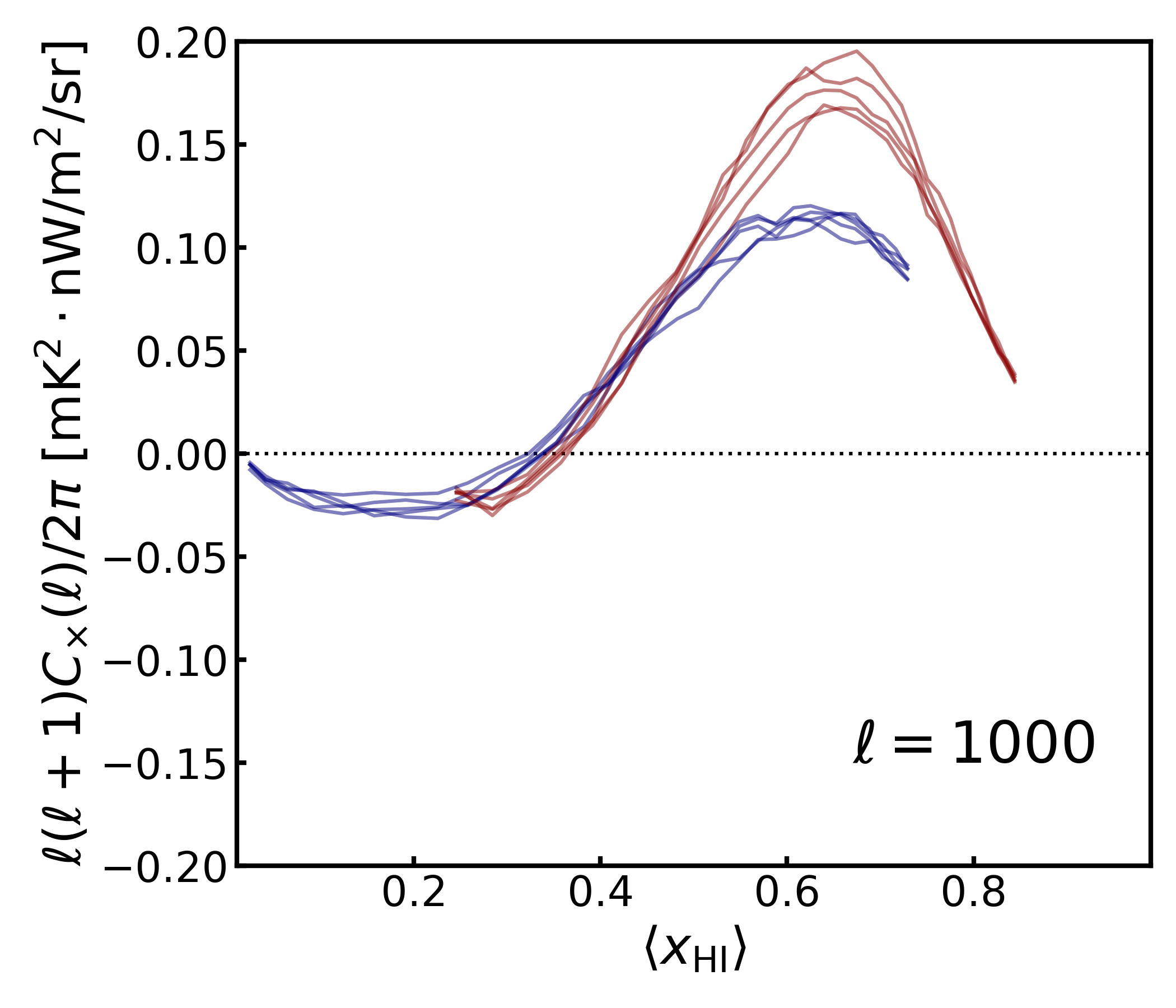}
 \caption{Left: the reionization timeline in our two reference models with momentum-driven vs energy-driven supernova feedback, in comparison with the observational constraints from the CMB and the dark pixel fraction in the Ly$\alpha$/Ly$\beta$ forest \cite[68\% and 95\% levels shown by the gray shaded bands;][]{Mason2019history,Whitler2020}. Middle: the evolution of the cross-power spectrum evaluated at $\ell=200$ with the IGM neutral fraction in the two reference models with different reionization histories. Results are shown only for the redshift range $6.5 < z < 9.5$ (as marked in the left panel). Since reionization in general occurs later in the case of energy-driven feedback, there is an offset in the plotted range of neutral fraction. Nevertheless, the zero-crossing of the cross-power spectrum occurs around the same neutral fraction of $\sim$40\% in both cases. For each case, results of 5 random realizations of the full EoR simulation are shown to give a sense of the sample variance. Right: same as the middle panel but for $\ell=1000$. }
 \label{fig:cps_vs_xhi}
\end{figure*}

Clearly, the unfiltered 21\,cm signal shows an increasingly pronounced anti-correlation with the NIRB on large scales as reionization proceeds, even though at the very early stage the correlation is weak or even reversed. The strong anti-correlation especially toward late stages of reionization is mainly a result of the opposite IGM phases traced by the two signals, whereas the overall stronger cross-correlation of the NIRB with lower redshift 21\,cm signal is because NIRB fluctuations are more dominated by lower redshift sources (see the bottom panel of Figure~\ref{fig:lightcones}). On the other hand, the overall weaker or even reversed correlation at early stages is associated with the smaller contribution of high-redshift sources to the NIRB and the different ways 21\,cm fluctuations are sourced, e.g., by spin temperature rather than ionized fraction fluctuations. Meanwhile, the cross-correlation signal turns over on progressively larger scales as the ionized bubbles grow, as in the case of the three-dimensional 21\,cm--galaxy cross-correlation \citep{Lidz2009,Park2014,Hutter2023,Moriwaki2024}.

For the filtered 21\,cm signal shown in the middle panels of Figure~\ref{fig:ccc}, as one would expect from Equation~(\ref{eq:TI_vanish}), the filtered cross-correlation signal vanishes at the ensemble average level. So while our simulation estimates based on averaging five realizations are not identically zero, the mean is consistent with zero as indicated by the shaded bands shown. From the right panels, the simple trick of squaring the filtered 21\,cm field as discussed in Section~\ref{sec:models:squared_cross} works as expected, yielding a non-vanishing signal that is highly informative when cross-correlated with the NIRB. In terms of scale dependence, the 21\,cm$^2$--NIRB cross-correlation gets stronger on larger scales, indicating the large-scale environments oppositely traced by these signals. In terms of time evolution, the cross-correlation starts mildly positive (reaching a correlation coefficient of $\gtrsim 0.4$) at the early stage of reionization, gradually weakens to a zero crossing point as reionization proceeds, and then grows increasingly negative (reaching a correlation coefficient of $\lesssim -0.6$) toward the late stage of reionization (see the left panel of Figure~\ref{fig:cps_vs_xhi} for the EoR history corresponding to these two modeled scenarios). Apart from the overall positive to negative transition of the cross-correlation coefficient, non-monotonic changes caused by the drop of correlation strength exist at the very early and very late stages (see also Figure~\ref{fig:cps_vs_xhi}). 

The way the cross-correlation signal traces the global reionization history is further illustrated in Figure~\ref{fig:cps_vs_xhi}, which shows the predicted reionization timeline and the 21\,cm$^2$--NIRB cross-power spectrum as a function of the IGM neutral fraction in the reference models with momentum-driven vs. energy-driven supernova feedback. As shown in the left panel, the two models represent, by construction, a plausible range of the IGM neutral fraction evolution in agreement with that inferred from combined measurements of the CMB optical depth and the dark pixel fraction in the Ly$\alpha$/Ly$\beta$ forest \citep{Mason2019history,Whitler2020}. From the middle and right panels, the co-evolution of the neutral fraction and the cross-correlation signal is clear on different scales. Interestingly, despite the very different reionization timelines, in both cases (and across all random realizations) the zero-crossing of the cross-correlation signal occurs around a neutral fraction of about 40\%. Meanwhile, the positive peak and the negative trough in both cases also occur around the same neutral fractions of about 60\% and 20\%, respectively, though the latter is subject to an increased sample variance. This correspondence indicates that characterizing the evolution of this 21\,cm$^2$--NIRB cross-correlation observationally provides a promising way to constrain the reionization timeline. While the 21\,cm power spectrum alone, which normally peaks near the reionization midpoint, can similary constrain the reionization history, the cross-correlation with the NIRB offers complementary information and serves as an independent check less susceptible to foreground contamination. 

\begin{figure*}[!ht]
 \centering
 \includegraphics[width=0.325\textwidth]{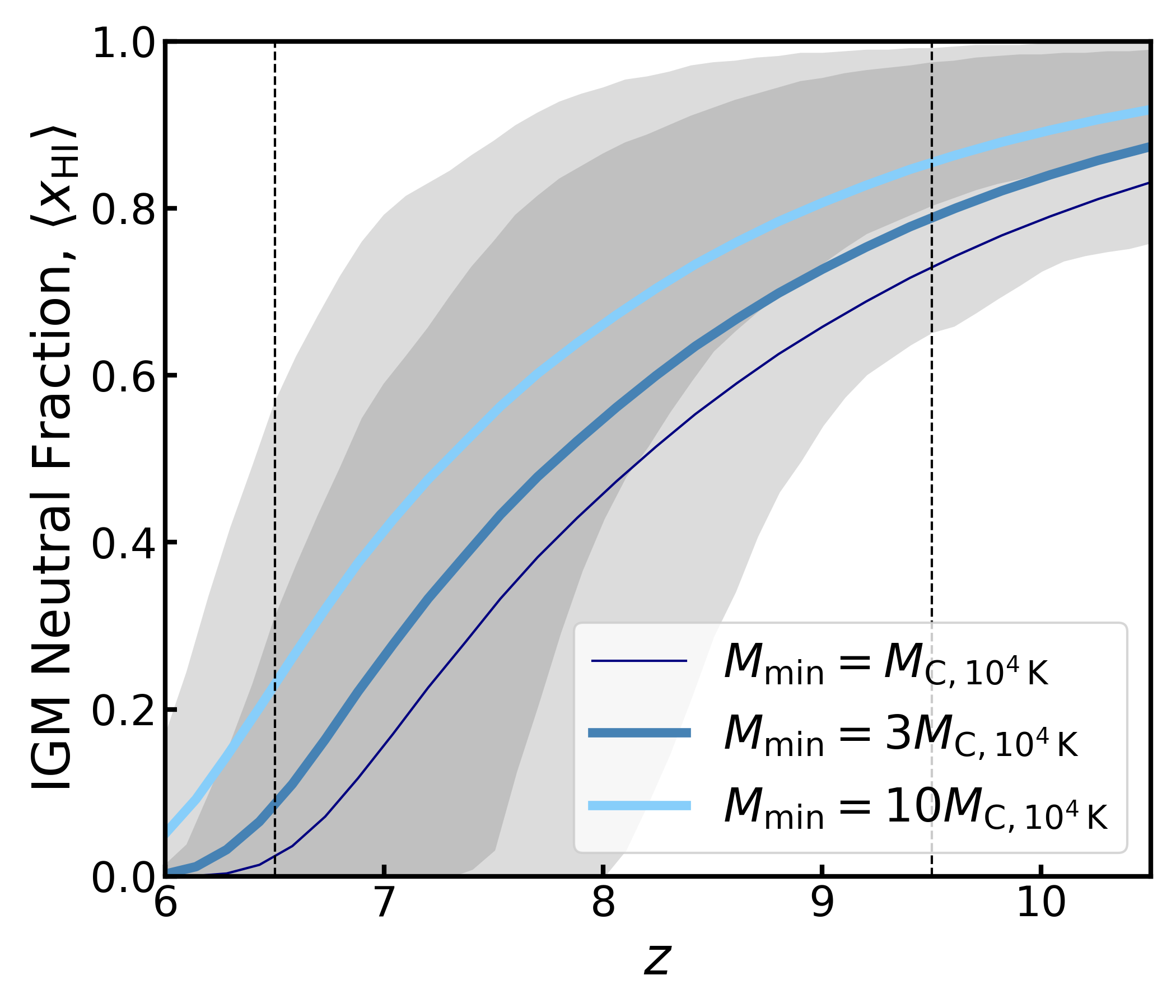}
 \includegraphics[width=0.325\textwidth]{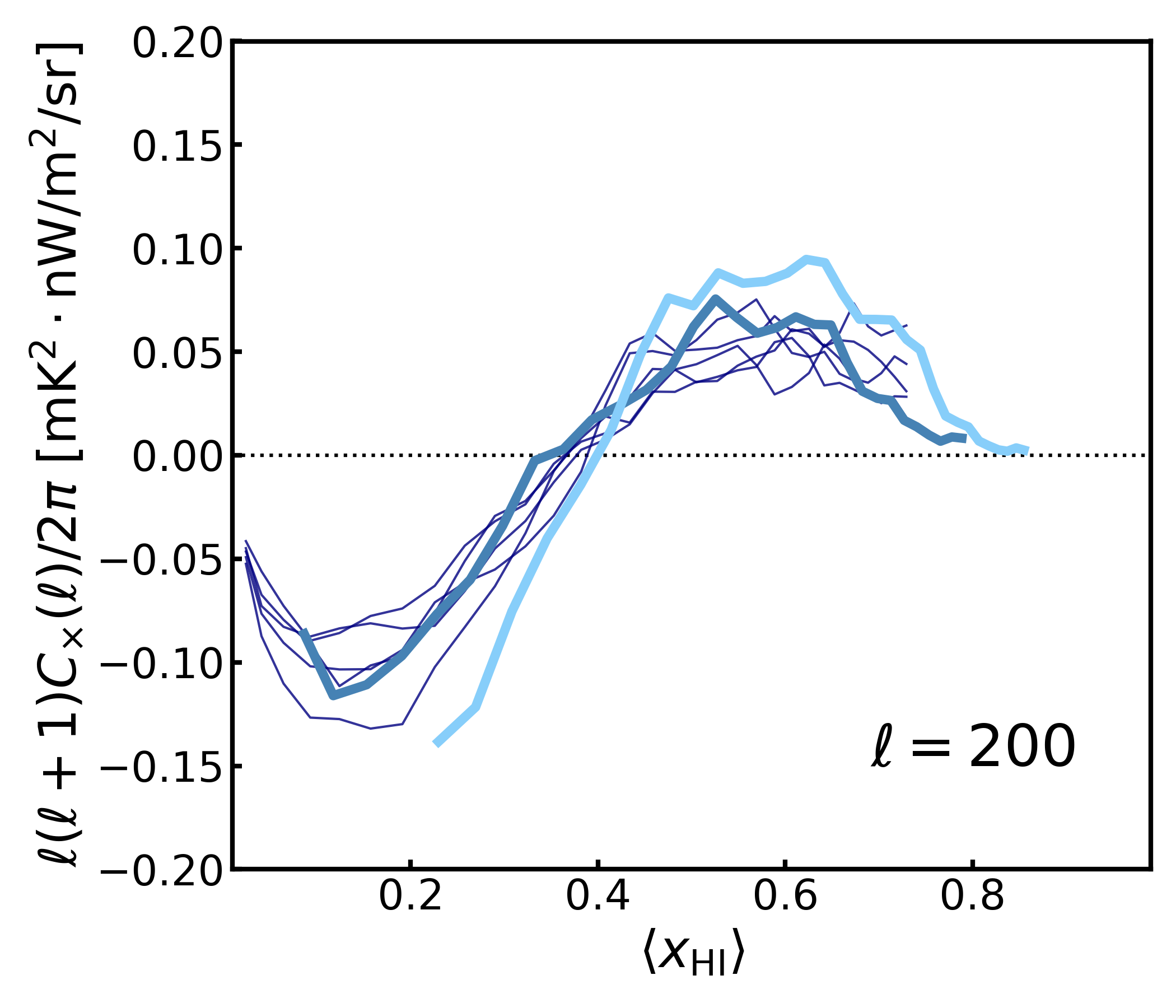}
 \includegraphics[width=0.325\textwidth]{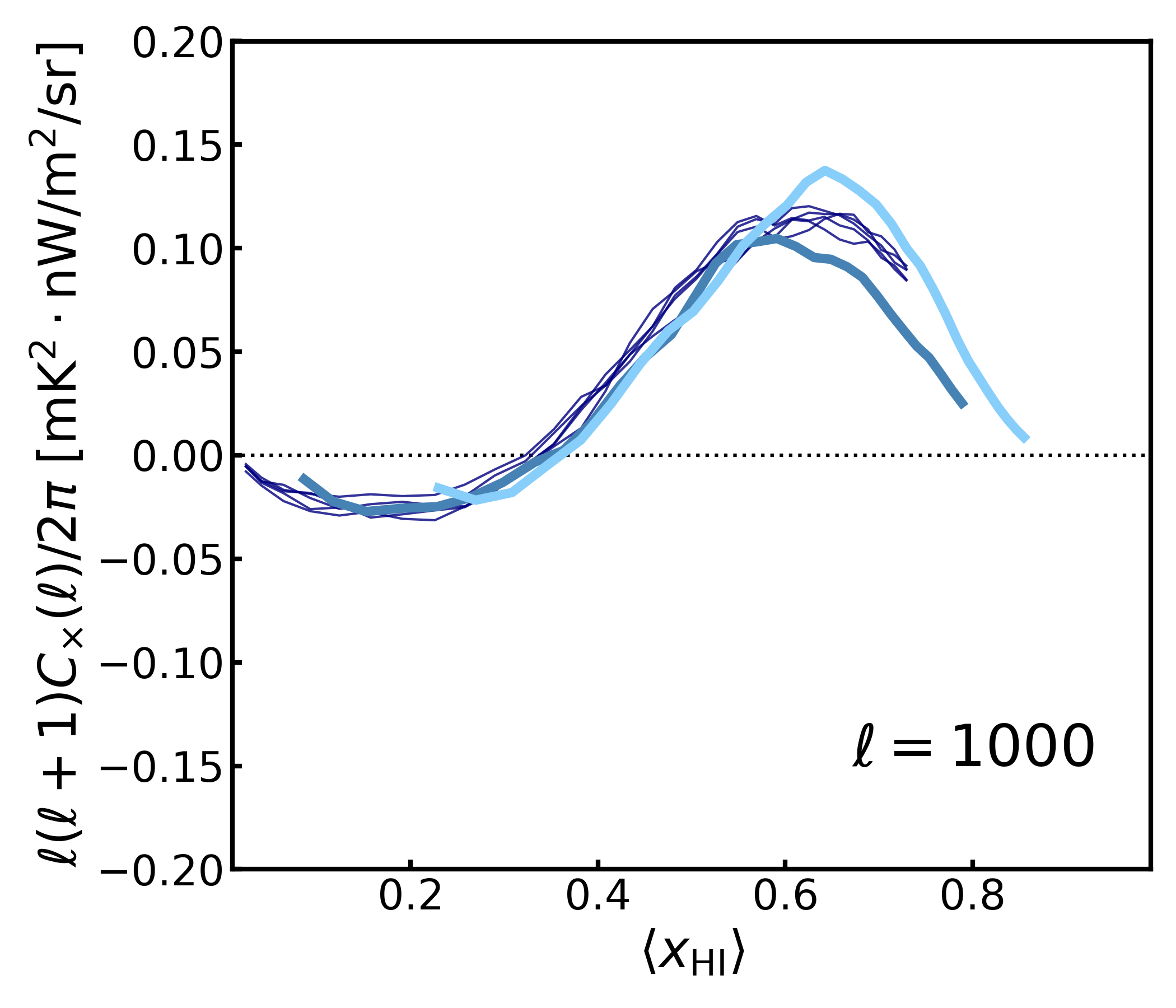}
 \caption{Same as Figure~\ref{fig:cps_vs_xhi}, but for different minimum halo mass for star formation $M_\mathrm{min}$ in the momentum-driven supernova feedback scenario. The two additional model variants assume $M_\mathrm{min}$ to be 3 and 10 times higher, respectively, than the fiducial value that equals to the atomic cooling threshold corresponding to a virial temperature $T_\mathrm{vir}=10^4\,$K. Although the reionization timeline is overall delayed in these two scenarios, the IGM neutral fraction at which the cross-power spectrum changes sign remains roughly the same ($\langle x_\mathrm{HI} \rangle \sim 0.4$).}
 \label{fig:cps_vs_xhi_mmin}
\end{figure*}

To better understand how robustly the reionization timeline is traced by the 21\,cm$^2$--NIRB cross-correlation, we examine another set of model variants for which the minimum halo mass for star formation, $M_\mathrm{min}$, is varied whereas the supernova feedback mode is fixed to be momentum-driven. In our fiducial models, $M_\mathrm{min}$ is set to be the mass threshold for efficient cooling by atomic hydrogen, $M_\mathrm{C,10^4\,K}$, which corresponds to a virial temperature of $T_\mathrm{vir} \sim 10^4\,$K. For the model variants, we increase $M_\mathrm{min}$ by a factor of 3 and 10, respectively, thereby delaying the overall reionization timeline. As an alternative way of modifying the EoR history without varying the feedback mode, this allows us to test whether diagnostics like the zero-crossing point can be applied to different reionization scenarios. As shown in Figure~\ref{fig:cps_vs_xhi_mmin}, the IGM neutral fraction $\langle x_\mathrm{HI} \rangle$ at which the cross-power spectrum changes sign remains roughly the same as $M_\mathrm{min}$ varies from $M_\mathrm{C,10^4\,K}$ ($\sim10^8\,M_{\odot}$ at $z=10$) to $10M_\mathrm{C,10^4\,K}$. Therefore, the fact that the zero-crossing point is insensitive to the feedback mode or $M_\mathrm{min}$ suggests that it can be generally leveraged to indicate a characteristic stage of reionization with $\langle x_\mathrm{HI} \rangle \sim 0.4$. 

\subsection{Detectability of the 21 cm$^2$--NIRB Cross-correlation with SKA and SPHEREx} \label{sec:results:detectability}

Now we have presented the approach to measure a non-vanishing cross-correlation signal between the 21 cm emission and the NIRB, which is shown to trace the time evolution of the IGM neutral fraction, it is of interest to understand whether this signal can be observationally detected. Here, we consider the synergy between SKA-Low and SPHEREx as an example case study to demonstrate the prospects for studying the reionization timeline with the 21\,cm$^2$--NIRB cross-correlation. 

To mimic 21\,cm foreground cleaning, in addition to the high-pass, sharp-$k_{\parallel}$ filter defined by Equation~(\ref{eq:sharpk}), we further take into account of the wedge effect in the $k_{\parallel}$--$k_{\perp}$ space, which arises from mode-mixing caused by the frequency-dependent instrumental response of 21\,cm interferometers. Specifically, we consider the contamination of modes within the wedge ($k_{\parallel} < m k_{\perp}$) and zero out them, with the foreground wedge slope given by \citep{LiuShaw2020}
\begin{equation}
m = k_{\parallel}/k_{\perp} = \frac{r(z) H(z) \sin\theta}{c(1+z)},
\label{eq:wedge}
\end{equation}
where $\theta$ is the beam angle. We emphasize that the foreground wedge effect is only considered for the detectability predictions shown in this section, but not the results in previous sections. As we demonstrate in Appendix~\ref{sec:appendix}, because a large fraction of the foreground wedge is already removed by the sharp-$k$ filter,  adding the foreground wedge filter here only modestly affects the resulting 21\,cm$^2$--NIRB cross-correlation such that its behavior discussed in Section~\ref{sec:results:cross} remains robust.  

In order to prevent noise-dominated modes from spreading and dominating in the squared field, we further apply a Wiener filer to down-weight noisy modes as an additional level of filtering in both scenarios. The choice of Wiener filter is motivated by the 21 cm signal and noise power spectra calculated from our simulations and assumed survey specifications \citep{LiuTegmark2012}, namely
\begin{equation}
F_{w}(k_{\parallel}, k_{\perp}) = \frac{P_{T}(k_{\parallel}, k_{\perp})}{P_{T}(k_{\parallel}, k_{\perp}) + P_{N,T}(k_{\parallel}, k_{\perp})}~.
\label{eq:wiener}
\end{equation}

Following \citet{Santos_2011}, we assume that stations within a compact core form a uniform baseline density distribution in the $u$--$v$ plane, which allows us to derive the instrumental thermal noise power (assumed to be a constant) from the radiometer equation as
\begin{equation}
P_{N,T} = \frac{T^2_\mathrm{sys}}{\Delta \nu t_\mathrm{eff}} V_\mathrm{vox} = \chi^2 y \frac{ \pi \lambda^2_\mathrm{obs} D^2_\mathrm{max} T^2_\mathrm{sys}}{t A^2_e N^2_\mathrm{ant}}~,
\end{equation}
where $\chi$ is the comoving angular diameter distance and $y = d\chi / d\nu = \lambda_\mathrm{obs}(1+z)/H(z)$. The effective observing time (per resolution $du^2$ in the $u$--$v$ plane) relates to the total observing time $t$ by $t_\mathrm{eff}/t=n(|\mathbf{u}|) du^2= n(|\mathbf{u}|)/\Omega_\mathrm{FOV}$ and the voxel size is $V_\mathrm{vox} = \chi^2 y \Delta \nu \Omega_\mathrm{FOV}$, with the field of view per antenna beam $\Omega_\mathrm{FOV}=\lambda^2_\mathrm{obs}/A_e$ and $n(|\mathbf{u}|)=\lambda^2_\mathrm{obs} N^2_\mathrm{ant}/\pi D^2_\mathrm{max}$. We can express the noise power of the filtered 21\,cm signal as
\begin{equation}
P_{N,\mathcal{T}}(k_{\parallel}, k_{\perp}) = |W_\mathrm{fg}(k_\parallel-k_\mathrm{\parallel,fg}) F_w(k_{\parallel}, k_{\perp})|^2 P_{N,T}
\end{equation}
where $W_\mathrm{fg}$ also includes the filtering of the foreground wedge. Thus, the noise for the filtered-and-squared signal is
\begin{equation}
P_{N,\mathcal{A}}(0, k_{\perp}) = 2 \int d k'_{\parallel} \int \frac{k'_{\perp} d k'_{\perp}}{(2\pi)^3} \int d \phi  P_{N,\mathcal{T}}(-k'_{\parallel}, |\boldsymbol{k}_{\perp}-\boldsymbol{k}'_{\perp}|) P_{N,\mathcal{T}}(k'_{\parallel}, k'_{\perp}), 
\end{equation}
where $\phi$ is the angle between vectors $k_{\perp}$ and $k'_{\perp}$. 

\begin{figure*}[!ht]
 \centering
 \includegraphics[width=0.495\textwidth]{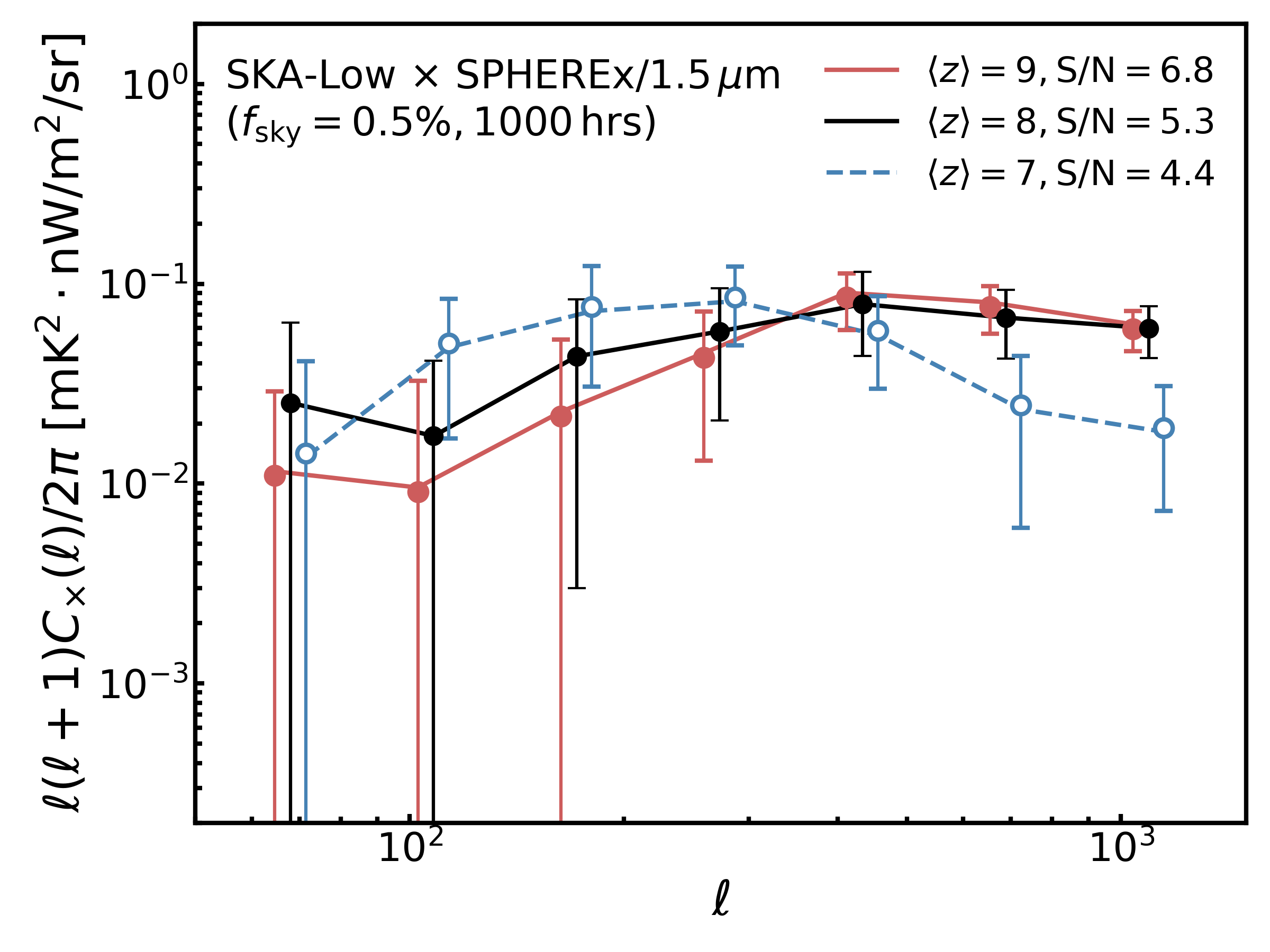}
 \includegraphics[width=0.495\textwidth]{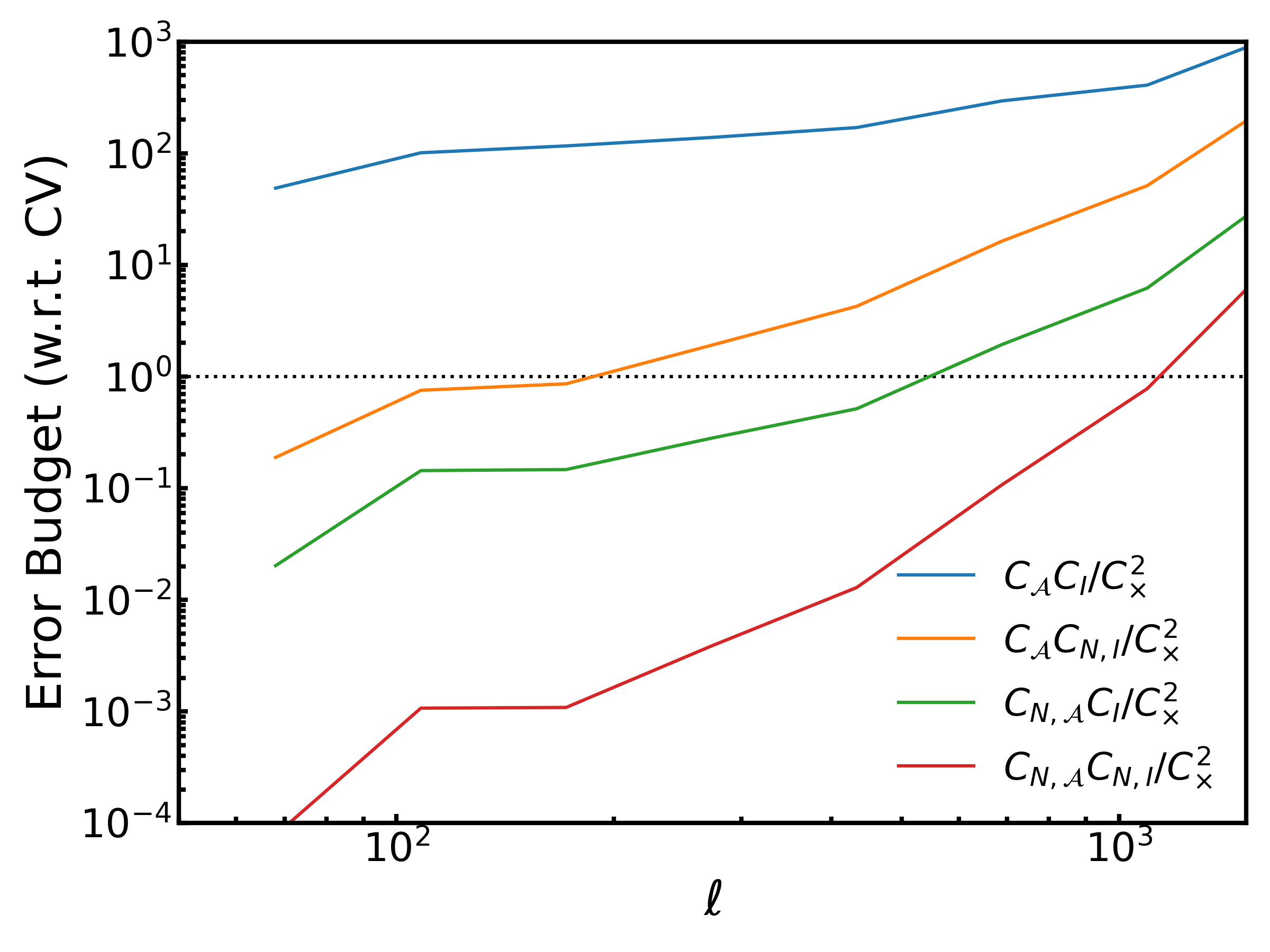}
 \caption{Left: the predicted detectability of the 21\,cm$^2$--NIRB cross-angular power spectrum, evaluated for 3 redshift bins ([6.5, 7.5], [7.5, 8.5], [8.5, 9.5]) assuming the instrumental specifications of SKA-Low and SPHEREx for the band centered at 1.5\,$\mu$m as detailed in the text. The absolute values are plotted as open circles and a dashed curve when the cross-angular power spectrum is negative, namely for the lowest redshift bin at $\langle z \rangle = 7$. The cross-correlation signal can be significantly detected at all 3 redshifts, with the total S/N summed over all $\ell$ bins slightly increasing with redshift due to the higher detectability on scales $\ell > 300$ at higher redshift. Right: a breakdown of the cross-power error budget term at $\langle z \rangle = 8$ in terms of the relative magnitude with respect to the sample variance term $C^2_{\times}$. Except on the smallest scales, the error budget is dominated by the cross term of the 21 cm and NIRB auto-power spectra.}
 \label{fig:detectability}
\end{figure*}

\begin{figure*}[!ht]
 \centering
 \includegraphics[width=0.495\textwidth]{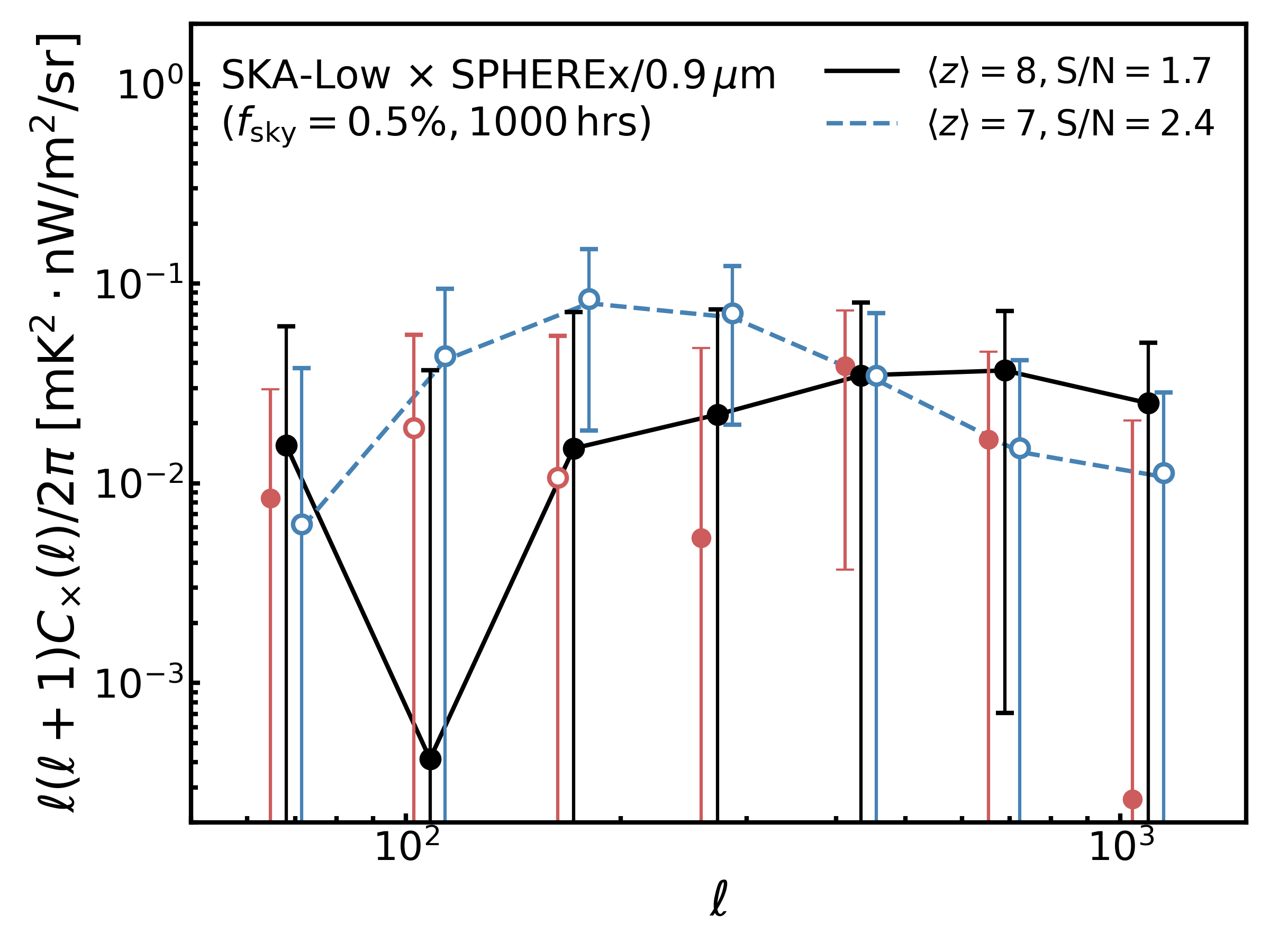}
 \includegraphics[width=0.495\textwidth]{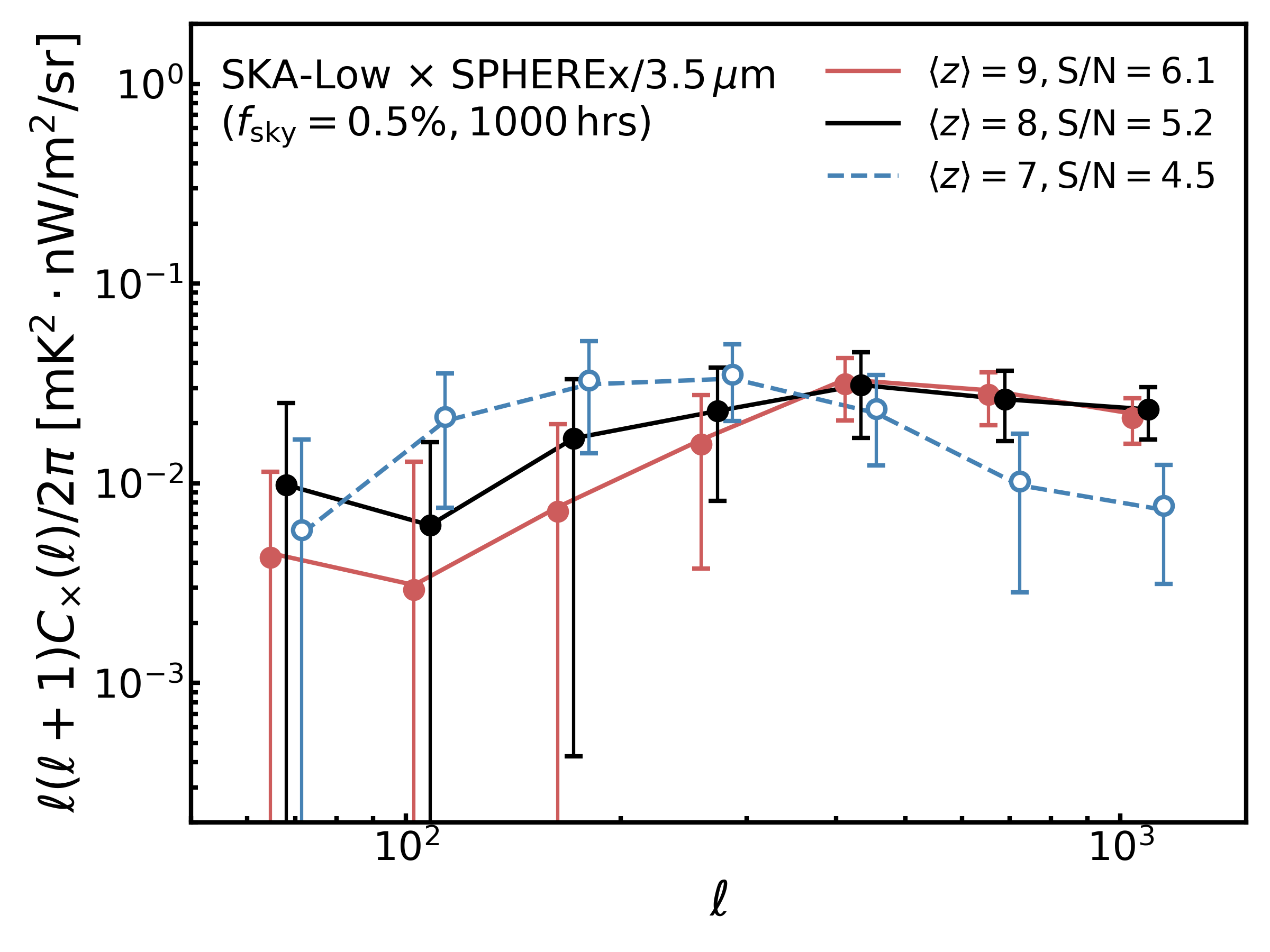}
 \caption{Same as the left panel of Figure~\ref{fig:detectability} but for two other different NIR bands of SPHEREx centered at 0.9\,$\mu$m (left) and 3.5\,$\mu$m (right), respectively. The absolute values are plotted as open circles and a dashed curve when the cross-angular power spectrum is negative. Since NIRB photons blueward of the rest-frame Ly$\alpha$ wavelength are attenuated by the IGM, the cross-power spectrum for the 0.9\,$\mu$m band diminishes and becomes noise-dominated at $z \gtrsim 8$.}
 \label{fig:detectability_others}
\end{figure*}

To clean 21\,cm foregrounds and predict the detectability of the cross-correlation signals (see Figures~\ref{fig:detectability} and \ref{fig:detectability_others}), we apply the sharp-$k$ filter defined by Equation~(\ref{eq:sharpk}) in the observing frame with a cutoff that corresponds to a physical scale of roughly 125\,cMpc/$h$ (or $k_\mathrm{\parallel,fg} \approx 0.05\,h$/cMpc), the foreground wedge filter as given by Equation~(\ref{eq:wedge}), along with the Wiener filter as defined by Equation~(\ref{eq:wiener}) using mock 21 cm signals and the SKA-Low instrument noise determined as follows. For $P_{N,T}$, we take $t = 1000\,$hr, $D_\mathrm{max} = 1000\,$m, $A_e = \pi(35\,\mathrm{m}/2)^2 = 962\,\mathrm{m^2}$, and $N_\mathrm{ant} = 512$. The system temperature $T_\mathrm{sys}$ scales as the sky temperature $T_\mathrm{sky}=60\left(\nu/300\,\mathrm{MHz}\right)^{-2.55}\,$K \citep{Thompson_2017} as $T_\mathrm{sys} = T_\mathrm{rcvr} + T_\mathrm{sky} = 1.1T_\mathrm{sky} + T_\mathrm{inst} \approx 225\,\mathrm{K}$ at 200\,MHz for $T_\mathrm{inst} = 40$\,K. To estimate uncertainties in the NIRB signal, we first set the instrument noise level according to the pixel size and the surface brightness sensitivity of the SPHEREx deep fields in the 1.5\,$\mu$m band\footnote{See the public product for the surface brightness sensitivity of SPHEREx available at \url{https://github.com/SPHEREx/Public-products/blob/master/Surface_Brightness_v28_base_cbe.txt}.}, namely $C_{N,I} = \sigma^2_\mathrm{pix} \Omega_\mathrm{pix}$. When estimating the uncertainty contributed by the NIRB auto-power spectrum, $C_{I}$, we enlarge the value of the contribution from $z>5$ sources that LIMFAST simulates by a factor of 10. This is to approximate the residual contribution from NIRB fluctuations and point sources at lower redshift that cannot be removed by masking or other component separation techniques \citep{Feng2019}. We leave a detailed investigation of the NIRB foreground modeling and subtraction to future work. 

Figure~\ref{fig:detectability} shows our predicted detectability of the 21\,cm$^2$--NIRB cross-angular power spectrum, together with a breakdown of its error budget, for the instrumental specifications of SKA-Low and SPHEREx described above. Similar to that shown in Figure~\ref{fig:ccc}, the cross power changes sign from positive at high redshift to negative at low redshift, with the peak moving toward larger scales (smaller $\ell$). Our sensitivity analysis suggests that the cross-correlation between a 1000-hour SKA-Low survey and the SPHEREx deep fields with a sky coverage of $f_\mathrm{sky} \approx 0.005$ can be detected in all of the 3 redshift bins considered at $6.5 < z < 9.5$, with a total $\mathrm{S/N} \gtrsim 5$ when summed over all scales. Due to the shift of the peak of the cross power spectrum, fluctuations on large, degree scales ($\ell \sim 200$) are more easily detectable at lower redshift, whereas small, $10{\arcmin}$-scale ($\ell \sim 1000$) fluctuations are more easily detectable at higher redshift. The breakdown of error budget shown in the right panel of Figure~\ref{fig:detectability} implies the cross term of the 21\,cm and NIRB auto-power spectra as the dominant source of uncertainty when measuring the cross-correlation. 

It is important to note that in Figure~\ref{fig:detectability} we only showcase the detectability of the cross-correlation for the 1.5\,$\mu$m broadband as an example.  In practice, cross-correlations can be performed at other SPHEREx wavelengths either with or without binning the native spectral channels into broadbands (Figure~\ref{fig:NIRB}). A multi-wavelength study can substantially boost the net detectability of the cross-correlation and unlock additional physical information about the ionizing sources \citep[e.g.,][]{Mirocha2022}. To illustrate the wavelength dependence of the cross-correlation signal, we show in Figure~\ref{fig:detectability_others} two more examples for the predicted detectability of the cross-power spectrum for broadbands centered at 0.9 and 3.5\,$\mu$m, respectively. As shown in the left panel, since NIRB photons blueward of the rest-frame Ly$\alpha$ wavelength are attenuated by the intervening neutral IGM, the cross-power spectrum for the 0.9\,$\mu$m band diminishes and becomes noise-dominated at $z \gtrsim 8$. On the other hand, as shown in the right panel, NIRB photons in the 3.5\,$\mu$m band have rest-frame wavelengths much longer than Ly$\alpha$ at the redshifts of interest and thus are not subject to this attenuation effect. The predicted cross-power spectrum and its detectability in this case are comparable to the results shown in Figure~\ref{fig:detectability}. 

Finally, we also note that, in addition to SPHEREx, the Looking at Infrared Background Radiation Anisotropies with Euclid (LIBRAE) survey of the Euclid mission \citep{Kashlinsky2015,Kashlinsky2019} plans to measure NIRB fluctuations in three colors (Y, J, and H) using Euclid's imaging of both the Wide ($\sim15,000\,\mathrm{deg^2}$) and Deep ($\sim40\,\mathrm{deg^2}$ in total) survey regions. The large sky coverage, along with the high sensitivity and angular resolution, of the LIBRAE survey make it another promising option for exploring the 21\,cm$^2$--NIRB cross-correlation that we propose.

\section{Discussion} \label{sec:discussion}

\subsection{The Physical Origin of the Sign Change} \label{sec:discussion:sign_change}

While the positive-to-negative transition of the 21 cm$^2$--NIRB cross-correlation has clearly proved itself to be a promising estimator for the reionization timeline, it remains to be seen what exactly causes such a transition. As we show in Sections~\ref{sec:models} and \ref{sec:results}, the unfiltered high-$k_{\parallel}$ 21\,cm power spectrum can be coupled with NIRB fluctuations through the (projected) cross-bispectrum, which leads to a non-vanishing cross-correlation measurable as a two-point statistic. Therefore, to better understand the transition, it can be instructive to inspect the coupling between 21\,cm and NIRB signals for an alternative yet simple estimator, such as the position-dependent 21\,cm power spectrum. 

\begin{figure*}[!ht]
 \centering
 \includegraphics[width=0.495\textwidth]{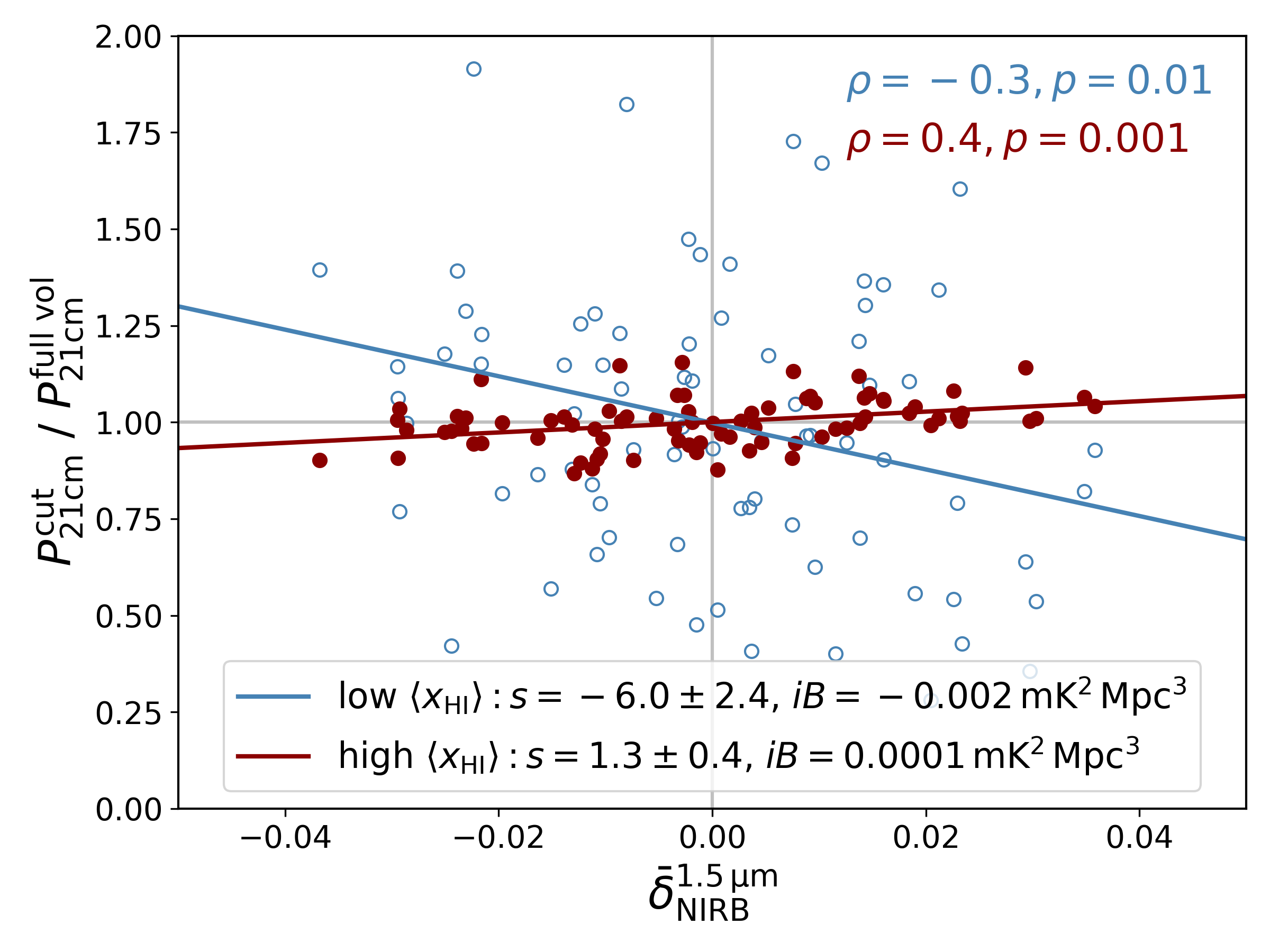}
 \includegraphics[width=0.495\textwidth]{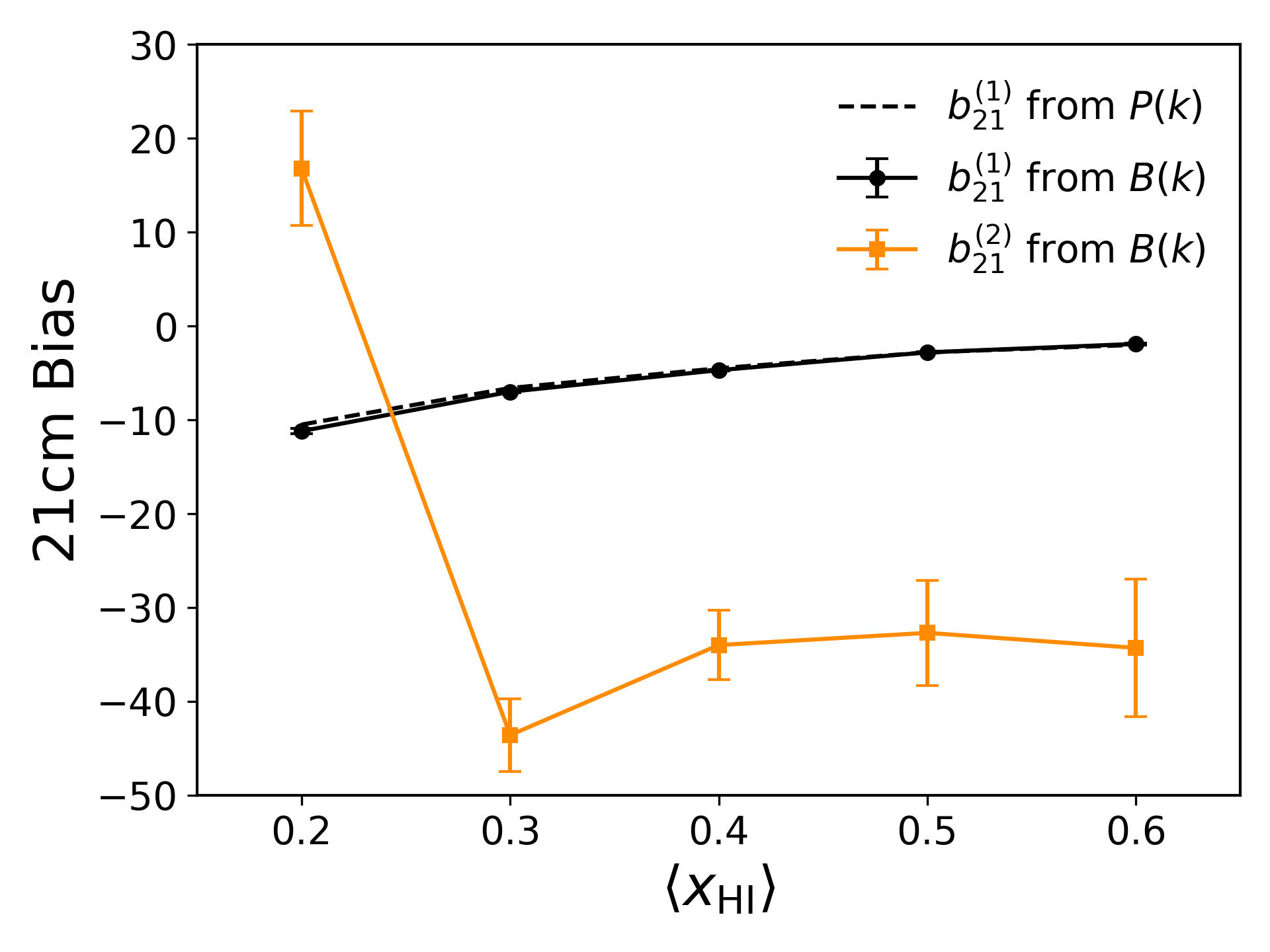}
 \caption{Left: correlation between the position-dependent, spherically-averaged 21 cm power spectrum and the local NIRB overdensity in two different regimes of IGM neutral fraction representing the early (``high $\langle x_\mathrm{HI} \rangle$'') and the late (``low $\langle x_\mathrm{HI} \rangle$'') stages of reionization, respectively. For each case, the Pearson correlation coefficient, $\rho$, and the associated p-value of the data points are displayed along with the best-fit linear relation. As indicated by the best-fit slopes shown in the legend, the sign of the integrated cross-bispectrum changes as reionization proceeds: in the early stages of reionization the large-scale NIRB hot-spots initially have a larger 21\,cm power spectrum than average, while NIRB hot-spots show  reduced 21\,cm power at later stages of reionization. The NIRB includes contributions from $6 < z < 15$, while the 21\,cm field is calculated over a much narrower range of redshift, which weakens and adds scatter around the correlation. Right: the $\langle x_\mathrm{HI} \rangle$ dependence of the first and second order 21\,cm bias factors inferred from fitting a simple bias expansion model to the 21\,cm power spectrum and bispectrum measured from our simulated data. The error bars denote 1-sigma uncertainties in the inferred bias factors. The rapid increase of $b^{(2)}_{T}$ from a negative value to a value greater than $|b^{(1)}_{T}|$ in the late stage of reionization provides a plausible explanation for the sign change observed for the 21\,cm$^2$--NIRB cross-correlation signal, as described by Equation~(\ref{eq:sign_change}).}
 \label{fig:ib_and_bias}
\end{figure*}

Here, we first validate our interpretation of the transition by calculating the position-dependent 21 cm power spectrum, whose correlation with the local mean density perturbation has been shown to be equivalent to an integral of the bispectrum \citep[see e.g.,][]{Chiang2014}, and investigating how it correlates with the local NIRB overdensity. The left panel of Figure~\ref{fig:ib_and_bias} summarizes the results of this analysis performed on mock data generated in the momentum-driven case. Each simulated light cone is divided into $N_\mathrm{cut}=16$ cutouts of equal angular size (3 deg$^2$) and a power spectrum $P^\mathrm{cut}_\mathrm{21\,cm}(k)$ is calculated for each cutout at $k = 0.1\,h/\mathrm{Mpc}$. We then plot the ratio of $P^\mathrm{cut}_\mathrm{21\,cm}(k)$ to the power spectrum evaluated for the full volume, $P^\mathrm{full\,vol}_\mathrm{21\,cm}(k)$, as a function of the mean NIRB overdensity for each cutout. Since the NIRB includes contributions from $6 < z < 15$, whereas the 21\,cm field is calculated over a much narrower range of redshift, a large scatter exists and weakens the correlation. Nevertheless, as indicated by the best-fit slopes, in the early stage of reionization ($8<z<9$) characterized by a high IGM neutral fraction, the position-dependent 21 cm power spectrum positively correlates with the NIRB overdensity, whereas in the late stage of reionization ($6<z<7$) characterized by a low IGM neutral fraction, the two quantities becomes anti-correlated. Such a transition is also encoded by the sign of the integrated bispectrum defined by \citep{Chiang2014}
\begin{equation}
iB(\boldsymbol{k}) = \frac{1}{N_\mathrm{cut}} \sum^{N_\mathrm{cut}}_{i=1} P(\boldsymbol{k}, \boldsymbol{r}_{L,i}) \bar{\delta}(\boldsymbol{r}_{L,i}),
\end{equation}
where $\bar{\delta}(\boldsymbol{r}_{L})$ is the local overdensity (taken to be the NIRB overdensity here) at position $\boldsymbol{r}$ averaged over some dimension $L$ and $N_\mathrm{cut}$ is the number of cutouts the total volume is divided into. Thus, as reionization progresses from early to late stages, 21 cm fluctuations change from being enhanced to being suppressed in overdense regions traced by NIRB emission (and vice versa). This may naturally explain the sign change in the 21 cm$^2$--NIRB cross-correlation, indicating that the position-dependent 21\,cm power spectrum (correlated with the NIRB) could be an alternate estimator for the EoR timeline. We should note, though, that given how modes probing different scales are weighted differently by these two estimators, the exact way they connect to each other is not obvious, which is an interesting matter for future work to address. 

For a slightly more quantitative understanding of what drives the sign change, it is useful to consider the following bias expansion (up to the second order) of the cross-bispectrum, $B_{T,T,I}$, which relates to the 21 cm$^2$--NIRB cross-correlation of interest by the projection given by Equation~(\ref{eq:bs_proj}), 
\begin{align}
B_{T,T,I}(k_1, k_2, k_3) \approx\ & 2 (b^{(1)}_{T})^2 b^{(1)}_{I} \left[ F_2 (k_1, k_2) P_\mathrm{lin}(k_1) P_\mathrm{lin}(k_2) + F_2 (k_1, k_3) P_\mathrm{lin}(k_1) P_\mathrm{lin}(k_3) + F_2 (k_2, k_3) P_\mathrm{lin}(k_2) P_\mathrm{lin}(k_3) \right] \nonumber \\
\ & b^{(2)}_{T} b^{(1)}_{T} b^{(1)}_{I} \left[ P_\mathrm{lin}(k_1) P_\mathrm{lin}(k_3) + P_\mathrm{lin}(k_2) P_\mathrm{lin}(k_3) \right] + b^{(2)}_{I} (b^{(1)}_{T})^2 \left[ P_\mathrm{lin}(k_1) P_\mathrm{lin}(k_2) \right],
\label{eq:cbs}
\end{align}
where
\begin{equation}
F_2(k_1, k_2) = \frac{5}{7} + \frac{1}{2} \cos \theta_{12} \left( \frac{k_1}{k_2} + \frac{k_2}{k_1} \right) + \frac{2}{7} (\cos \theta_{12})^2
\end{equation}
and
\begin{equation}
\cos \theta_{12} = \frac{k_3^2 - k_1^2 - k_2^2}{2 k_1 k_2}.
\end{equation}
Assuming a squeezed, isosceles triangle configuration ($k_1 = k_2 \gg k_3$), from which the contribution to $P^\mathrm{2D}_{\mathcal{A},I}(k_{\perp})$ or $C_{\mathcal{A},I}(\ell)$ is expected to dominate (given that the signal mainly measures the coupling between large-scale transverse modes of the NIRB and small-scale LOS modes of the 21 cm line; Equation~(\ref{eq:bs_proj})), and dropping the last term of Equation~(\ref{eq:cbs}) for $b^{(2)}_{I} \sim 0$, we can express the ratio of the first two terms as
\begin{equation}
R = \frac{2b^{(1)}_{T}}{b^{(2)}_{T}} \frac{F(k_1,k_1) P_\mathrm{lin}(k_1)^2 + 2 F(k_1,k_3) P_\mathrm{lin}(k_1) P_\mathrm{lin}(k_3)}{2 P_\mathrm{lin}(k_1) P_\mathrm{lin}(k_3)} \sim 2 b^{(1)}_{T} F(k_1,k_3) / b^{(2)}_{T} \sim b^{(1)}_{T} / b^{(2)}_{T}.
\label{eq:sign_change}
\end{equation}

Since $b^{(1)}_{T}$ evolves almost linearly with the IGM neutral fraction, $\langle x_\mathrm{HI} \rangle$, and becomes more and more negative as $\langle x_\mathrm{HI} \rangle$ decreases \citep[see e.g.,][which is verified with our own simulations]{Hoffmann2019}, for $B_{T,T,I}$ to have a sign change we must have $R$ reach $-1$ by driving a positive $b^{(2)}_{T}$ to exceed $|b^{(1)}_{T}|$ as $\langle x_\mathrm{HI} \rangle$ evolves from the early to the late stage of reionization. This is in line with the results in \citet{Hoffmann2019}, as well as what we find for $b^{(2)}_{T}$ by fitting the following simple bias expansions of the 21 cm power spectrum and bispectrum to our simulated 21 cm data
\begin{equation}
P_{T} \approx [b^{(1)}_{T}]^2 P_\mathrm{m}
\end{equation}
and
\begin{equation}
B_{T} \approx [b^{(1)}_{T}]^3 B_\mathrm{m} + [b^{(1)}_{T}]^2 b^{(2)}_{T} \left[ P_\mathrm{m,12} P_\mathrm{m,13} + P_\mathrm{m,21} P_\mathrm{m,23} + P_\mathrm{m,31} P_\mathrm{m,32} \right],
\end{equation}
where $P_\mathrm{m}$ and $B_\mathrm{m}$ are the power spectrum and bispectrum of matter overdensity. Even though the exact IGM neutral fractions where $b^{(2)}_{T}$ undergoes an abrupt change do not perfectly match, results based on our simple bias expansion as shown in the right panel of Figure~\ref{fig:ib_and_bias} imply that the sign change of the 21 cm$^2$--NIRB cross-correlation may be attributed to the change in the 21\,cm--matter overdensity correlation captured by the second order 21\,cm bias, given that the NIRB is a good tracer of matter overdensities. 

\subsection{Limitations and Possible Extensions of the Presented Framework}

While we updated the theoretical framework and sensitivity forecasts for the joint analysis between 21\,cm and cosmic NIRB signals, the presented framework is not free of limitations that need to be improved in future studies. First of all, the source modeling of the NIRB is still simplified and incomplete. In particular, LIMFAST only simulates the high-redshift contribution to the cosmic NIRB from sources down to $z = 5$ whereas it is well-known that sources at lower redshifts contribute the majority of the NIRB signal \citep{Feng2019,ChengChang2022}. We have employed a fairly crude treatment to enlarge the high-$z$ NIRB auto-power spectrum from LIMFAST by a factor 10 in order to account for the inseparable low-redshift contribution in our sensitivity forecasts, but a more accurate treatment based on self-consistent NIRB modeling down to much lower redshift (Mirocha et al. in prep.) will be desirable for more detailed investigations of the cross-correlation signal proposed here. Meanwhile, although the model of galaxy formation and evolution adopted is already an improvement compared to previous studies, it is similar to semi-empirical prescriptions invoking an SFE that barely evolves with time \citep[e.g.,][]{Mason2015,SF2016,Mirocha2017,Tacchella2018}. Some models motivated by recent observations from the James Webb Space Telescope (JWST) suggest an enhanced SFE due to inefficient stellar feedback\footnote{Note that an enhanced SFE is just one possible explanation for the number density of bright galaxies observed at cosmic dawn by JWST \citep[e.g.,][]{Mirocha2023,Sun2023UVLF,Feldmann2024,Ferrara2024,Yung2024}.} in galaxies at cosmic dawn \citep{Inayoshi2022,Dekel2023,Harikane2023}, which can potentially impact observables of the EoR including the 21\,cm global signal and fluctuations \citep{Hassan2023,Libanore2024}. It is thus instructive for future work to enable higher flexibility in source modeling to better explore the possible impact of different physical conditions of early galaxies on the 21\,cm and NIRB signals. Furthermore, attenuation of the (intrinsic) stellar and nebular emission from galaxies by interstellar dust may have non-trivial effects on the observed NIRB spectrum even at $z>5$. In future studies, it would be useful to incorporate semi-empirical models of dust attenuation \cite[e.g.,][]{Finke2022,Mirocha2022} into our framework to more accurately simulate the NIRB. 

Regarding the 21 cm$^2$--NIRB cross-correlation, we should emphasize that our sensitivity estimates in Section~\ref{sec:results:detectability} still involve a number of simplifying assumptions. Our S/N estimation essentially neglects any 21\,cm foreground residual contributions to the predicted error bars, which add to the variance even if not correlated on average with the NIRB. Also, 21\,cm foreground mitigation strategies more sophisticated than the simple cases considered are likely needed. For the NIRB, while we include a simple estimation of the residual contamination from low-$z$ galaxies after masking by boosting the high-$z$ auto-power spectrum $C_{I}$ by a factor of 10, the residual might be higher in practice and reduce the correlation in addition to adding variance. Meanwhile, the low-$z$ residual of the NIRB can also be partially correlated with the residual 21\,cm foreground, thus further complicating the analysis. Finally, instead of assuming gaussian errors, the full covariance of the three-point statistics may be worth consideration. Notwithstanding these intricacies that future studies should investigate, our approach to obtain a non-vanishing cross-correlation signal is by no means limited to the 21\,cm--NIRB joint analysis. It is generally applicable to cross-correlations of 3D line IM data and other cosmological probes in 2D, such as the SZ effect and lensing of the CMB, the unresolved CXB, and photometric galaxy surveys \citep[see also][]{Zhu2018}. 

\section{Conclusions} \label{sec:conclusions}

In this paper, by forward modeling the cosmic 21\,cm signal and the NIRB associated with high-$z$ star-forming galaxies with the semi-numerical code LIMFAST, we perform an updated analysis of their cross-correlation as a probe of the EoR timeline. Our physically motivated, multi-tracer framework significantly improves over previous studies in that the model ingredients, especially our prescriptions for star formation, are carefully calibrated against the latest observations of the high-$z$ galaxy population and reionization. We then self-consistently calculate the 21 cm signal and the NIRB of high-$z$ origin ($z>6$). These semi-numerical simulations are also flexible and computationally efficient, allowing us to easily generate multiple realizations of cosmic reionization and the corresponding observables under different model assumptions. 

With this powerful modeling framework in hand, we focus particularly on developing a simple method to cross-correlate the 21 cm and NIRB signals with realistic considerations of the foreground contamination, which was largely neglected in previous studies. While the two signals generally trace opposite phases of the IGM during the EoR, we demonstrate that due to the mismatched Fourier space coverage after the removal of the foreground-contaminated long-wavelength modes of the 21 cm signal along the line of sight, a direct cross-correlation between the foreground-filtered 21 cm signal and the NIRB inevitably vanishes. To resolve this problem and recover the invaluable information offered by the cross-correlation, we square the foreground-filtered 21 cm signal and then cross-correlate it with the NIRB. This solution effectively defines a new two-point statistics that is equivalent to a projected cross-bispectrum, where the higher order statistics couples together transverse NIRB modes with the short-wavelength line-of-sight 21 cm modes that survive the foreground cleaning. 

The resulting 21 cm$^2$--NIRB cross-correlation signal is non-vanishing and serves as a promising probe for the timeline of cosmic reionization by closely tracing the IGM neutral fraction evolution. In the early stage of reionization with $\langle x_\mathrm{HI} \rangle > 0.5$, the 21 cm$^2$--NIRB cross-correlation is positive, reaching a correlation coefficient of $\gtrsim0.4$ on large ($\sim 1\,$deg) scales (assuming the $z < 6$ contributions to the NIRB are masked out or filtered away). As reionization progresses, the correlation continuously decreases, crosses zero near $\langle x_\mathrm{HI} \rangle \sim 0.4$, and then becomes increasingly negative toward the late stage of reionization, reaching a correlation coefficient of $\lesssim -0.6$ on large scales. This positive-to-negative transition of the 21 cm$^2$--NIRB cross-correlation, including the zero-crossing, establishes an easy-to-measure correspondence with the IGM neutral fraction evolution. Therefore, in addition to validating possible 21 cm detections, the sign and strength of this cross-correlation signal also offer a viable means to measure the reionization timeline. Using mock signals generated by LIMFAST, we further quantify the detectability of the 21 cm$^2$--NIRB cross-angular power spectrum at different reionization stages by a synergy between SKA-Low and SPHEREx. Under the assumption of a 1000-hour SKA-Low survey in overlap with the 200\,deg$^2$ SPHEREx deep fields, we find that the cross-power spectrum, including its sign change, can be significantly detected during most of the reionization history $6.5 < z < 9.5$. It is thus reasonable to expect measurements of the reionization timeline based on solid detections of this cross-correlation signal by these forthcoming facilities. 

In summary, the 21 cm$^2$--NIRB cross-correlation not only offers a simple solution to extract the complementary information about the EoR provided by the NIRB and the foreground contaminated 21 cm data, but also opens an window for understanding the detailed reionization timeline and the physics that shapes the connection between the IGM and ionizing sources on different scales. 

\section*{Acknowledgment}
We thank the anonymous reviewer for useful comments, as well as Nick Gnedin, Simon Foreman, Paul La Plante, Yin-Zhe Ma, Kana Moriwaki, Julian Mu\~{n}oz, Clinton Stevens, Naoki Yoshida, Rui Lan Zhang, and Meng Zhou for stimulating discussions. G.S. was supported by a CIERA Postdoctoral Fellowship and acknowledges the Kavli Institute for Theoretical Physics (KITP) where part of this work was done for their hospitality, which is supported in part by grant NSF PHY-2309135. A.L. acknowledges support from NASA ATP grant 80NSSC20K0497. T.-C.C. acknowledges support by NASA ROSES grant 21-ADAP21-0122. J.M. was supported by an appointment to the NASA Postdoctoral Program at the Jet Propulsion Laboratory/California Institute of Technology, administered by Oak Ridge Associated Universities under contract with NASA. Part of this work was done at Jet Propulsion Laboratory, California Institute of Technology, under a contract with the National Aeronautics and Space Administration (80NM0018D0004).  S.R.F. was supported by NASA through award 80NSSC22K0818 and by the National Science Foundation through award AST-2205900. \\

\begin{figure*}[!t]
 \centering
 \includegraphics[width=0.495\textwidth]{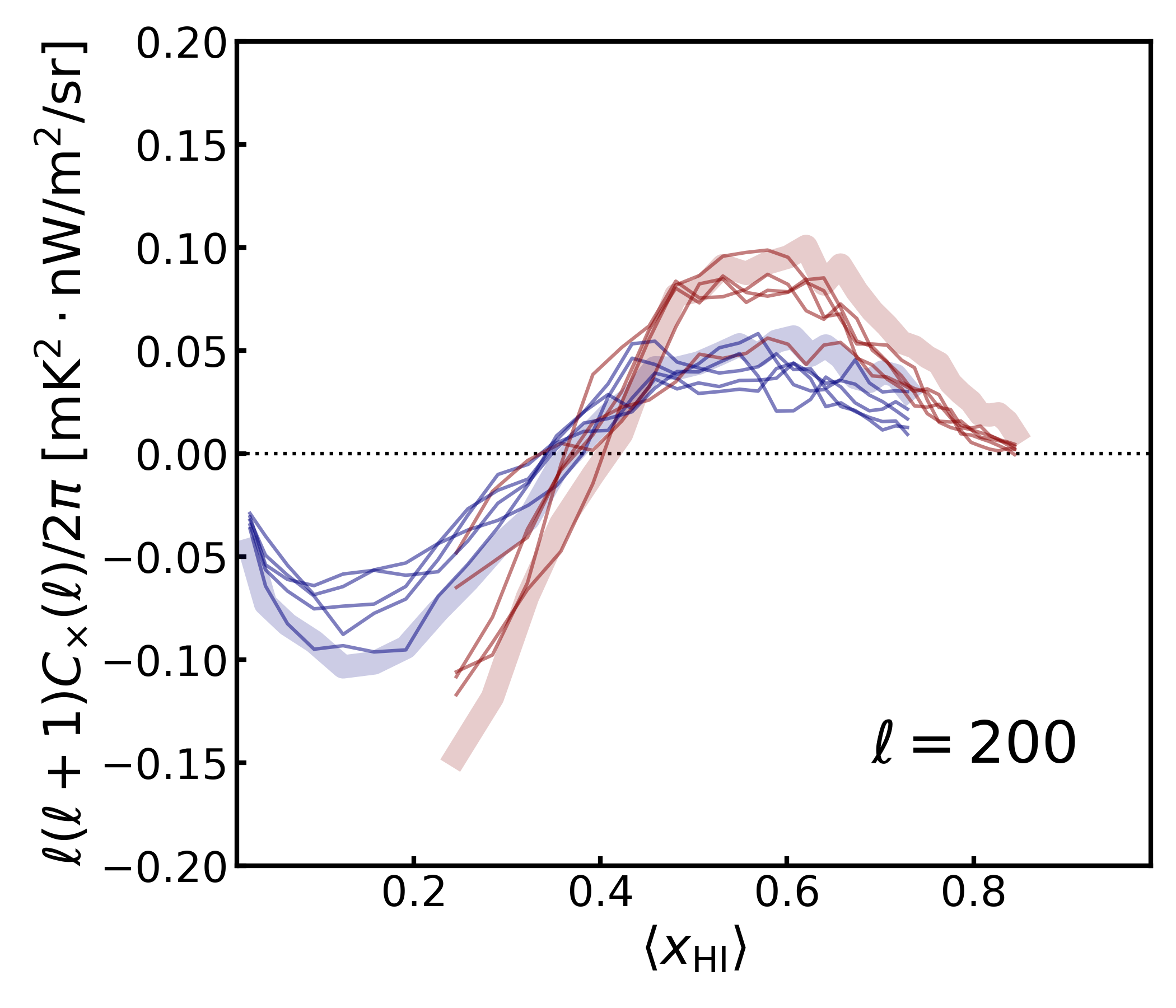}
 \includegraphics[width=0.495\textwidth]{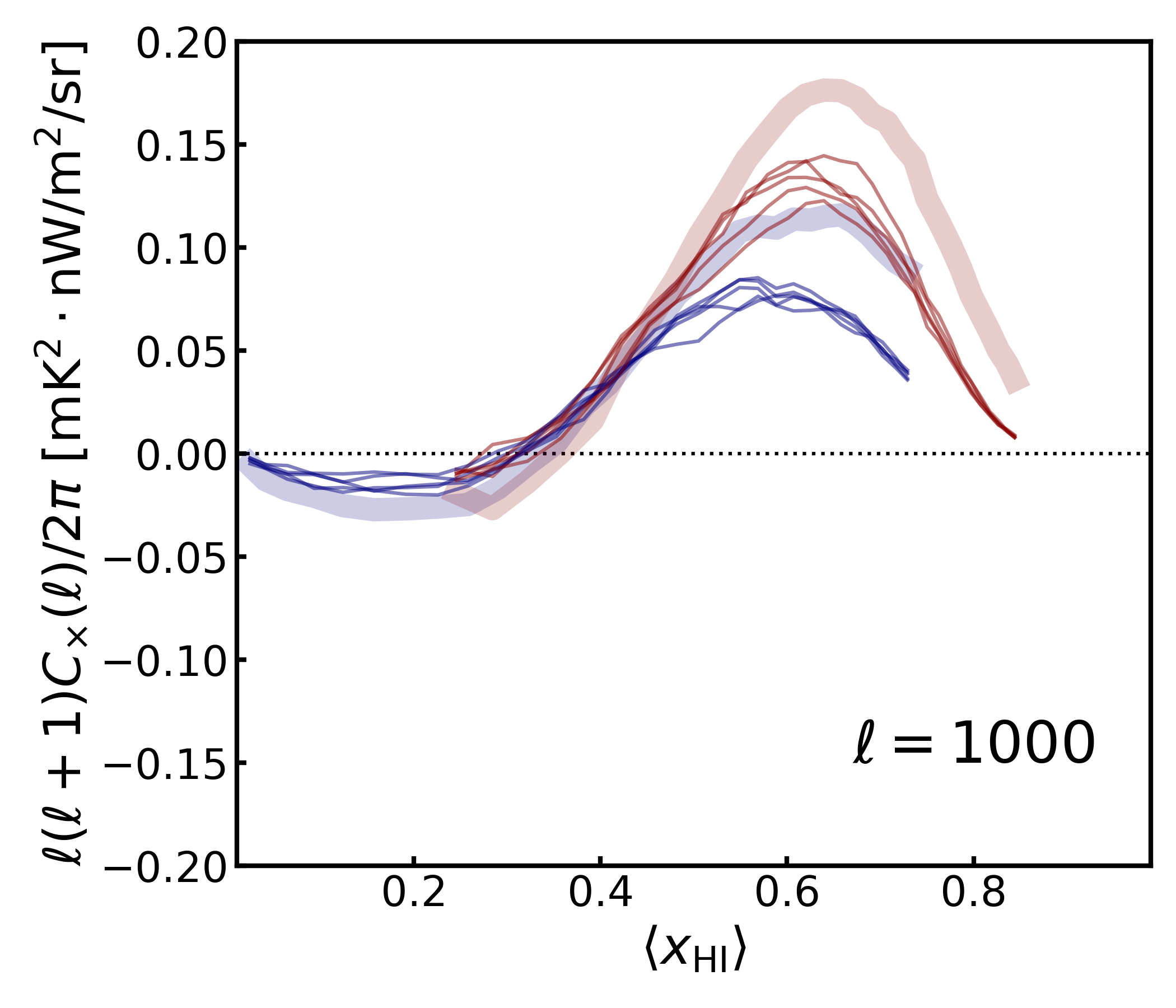}
 \caption{Same as the middle and right panels of Figure~\ref{fig:cps_vs_xhi} but with the 21\,cm foreground wedge removed in addition to the lowest $k_{\parallel}$ modes (the Wiener filter is still applied). For comparison, the median results in Figure~\ref{fig:cps_vs_xhi} without the wedge subtraction are overplotted as shaded bands. Foreground wedge removal modestly reduces the amplitude of the angular cross-power spectrum by no more than a few tens percent and shifts the zero-crossing point and the two extrema to IGM neutral fractions that are $\sim0.05$ lower compared to the case when only the high-pass, sharp-$k_{\parallel}$ filter is applied.}
 \label{fig:cps_vs_xhi_wedge}
\end{figure*}

\appendix

\section{Effects of Foreground Wedge Removal on the Cross-correlation} \label{sec:appendix}

To keep the discussion simple, in Section~\ref{sec:results:cross} we present the 21\,cm$^2$--NIRB cross-correlation signals without considering the foreground wedge effect. Filtering of modes in the foreground wedge is included in Section~\ref{sec:results:detectability}, however, to more reliably forecast the detectability of the cross-correlation signal. For comparison, we show in Figure~\ref{fig:cps_vs_xhi_wedge} the 21\,cm$^2$--NIRB cross-power spectrum vs. the IGM neutral fraction $\langle x_\mathrm{HI} \rangle$ similar to Figure~\ref{fig:cps_vs_xhi_wedge} but when the entire foreground wedge is filtered. Overall, the cross-power spectrum is only modestly affected by removing the foreground wedge (a large fraction of which is already removed by the sharp-$k$ filter) and the key features identified remain robust. 

\bibliography{paper3}{}
\bibliographystyle{aasjournal}

\end{document}